\documentclass[prab,amsmath,amssymb,reprint,longbibliography]{revtex4-1}

\pdfoutput=1

\usepackage[utf8]{inputenc}

\usepackage{xcolor,colortbl}
\usepackage{siunitx,graphicx,hyperref,siunitx}
\usepackage{algorithm}
\usepackage[noend]{algpseudocode}
\usepackage[caption=false]{subfig}
\usepackage{booktabs}

\begin{document}

\title{Multiobjective optimization of the dynamic aperture for SLS~2.0\\using surrogate models based on artificial neural networks}
\author{M. Kranj\v{c}evi\'{c}}
\email{marija.kranjcevic@psi.ch}
\author{B. Riemann}
\author{A. Adelmann}
\author{A. Streun}
\affiliation{Paul Scherrer Institut, CH-5232 Villigen PSI, Switzerland}
 
\begin{abstract}
Modern synchrotron light source storage rings, such as the Swiss Light Source upgrade (SLS~2.0), use multi-bend achromats in their arc segments to achieve unprecedented brilliance. This performance comes at the cost of increased focusing requirements, which in turn require stronger sextupole and higher-order multipole fields for compensation and lead to a considerable decrease in the dynamic aperture and/or energy acceptance. In this paper, to increase these two quantities, a multi-objective genetic algorithm (MOGA) is combined with a modified version of the well-known tracking code \texttt{tracy}. As a first approach, a massively parallel implementation of a MOGA is used. Compared to a manually obtained solution this approach yields very good results. However, it requires a long computation time. As a second approach, a surrogate model based on artificial neural networks is used in the optimization. This improves the computation time, but the results quality deteriorates. As a third approach, the surrogate model is re-trained during the optimization. This ensures a solution quality comparable to the one obtained with the first approach while also providing an order of magnitude speedup. Finally, good candidate solutions for SLS~2.0 are shown and further analyzed.
\end{abstract}

\maketitle

\section{Introduction}

The upgrade of the Swiss Light Source, called SLS~2.0, is scheduled for 2023--24. To increase the brilliance, the current 3rd generation storage ring will be replaced by one employing seven-bend achromats, including reverse bends and longitudinal gradient bends~\cite{PhysRevAccelBeams.22.021601}. 
The stronger focusing requirements need higher sextupole and higher-order multipole fields for chromatic compensation. 
This makes finding a reasonably large dynamic aperture (DA) for injection and an energy acceptance for a sufficient beam lifetime more challenging and more important. It can either be done indirectly, by computing and minimizing the dominant resonance driving terms~\cite{johan-sls2}, or directly, by computing and maximizing the DA and energy acceptance~\cite{ehrlichman,li-cheng}.

In this work the latter approach is used and a constrained multiobjective optimization problem is formulated (section~\ref{sec:opt-problem}). The search space comprising the strengths of sextupole families, as well as horizontal and vertical linear chromaticity is considered.
Similarly to the approach in~\cite{ehrlichman}, the objective functions are defined to maximize the transverse DAs at three different energies and to prevent the tune resonances from being crossed, thus maximizing the energy acceptance and beam lifetime. These figures of merit are computed using direct particle tracking with a modified version of the well-known tracking code \texttt{tracy}~\cite{tracy}.

Out of the many multiobjective optimization algorithms, particle swarm optimization~\cite{safranek}, differential evolution~\cite{husain,wu} and multiobjective genetic algorithms~\cite{ipac20,ehrlichman,gao,li-yang,hofler:13,yang2011,yang2009} have already been successfully applied to the problem of optimizing the DA. In this work a multiobjective genetic algorithm (MOGA) is chosen and further extended with constraint-handling methods 
(section~\ref{sec:moga}).

Previous work includes approaches that speed up the convergence of the multi-generation optimization method by, e.g., preselecting points that are likely to be good using $k$\nolinebreak-means clustering~\cite{li-cheng} or a \emph{surrogate model}~\cite{wang-2019,gaussian-process,gaussian-process-appl}, i.e., an approximation model which captures the significant properties of a given simulation model and is also very cheap to evaluate. In this work an artificial neural network (ANN) surrogate model is used for the optimization, in combination with a MOGA, similarly to~\cite{edelen}. 
Additionally, the solution quality is improved by re-training the surrogate model during the optimization.

First, the ANN surrogate model is built to approximate the objective functions (section~\ref{sec:building-surrogate-model}). In particular, good hyperparameters are determined and the surrogate model quality is shown. The surrogate model is then used for optimization and a new re-training procedure is devised (section~\ref{sec:optimizing-with-the-surrogate-model}). The run time and the solution quality of the new approach are compared with those of a massively parallel implementation of a MOGA coupled with \texttt{tracy}. The solution quality is determined in comparison with an existing manually obtained solution.

Finally, good candidate solutions are shown and further analyzed (section~\ref{sec:results}). In particular, the transverse DAs at the three considered energies are shown and compared with those of the manually obtained solution.

\section{Optimization problem\label{sec:opt-problem}}

\subsection{Dynamic aperture (DA)\label{sec:DA}}
The DA can be loosely defined as an area in the transverse phase space in which stable particle motion can occur.
To quantify the size of the DA the approach from~\cite{ehrlichman} together with the modifications from~\cite{ipac20} is adopted.
As the DA area is dependent on the local linear optics given by the Twiss parameters $\alpha$ and $\beta$ at the starting location of the particle tracking, the DA coordinates $r$ and $\theta$ in Floquet space are mapped to the coordinates for tracking via
\begin{equation}
    \begin{aligned}
    \begin{pmatrix} x \\ x^\prime \end{pmatrix} &= 
    \begin{pmatrix} \beta \\ -\alpha \end{pmatrix}_x \frac{r \cos \theta_k}{\sqrt{\beta_x}}, \\
    \label{eq:floquet}
    \begin{pmatrix} y \\ y^\prime \end{pmatrix} &= 
    \begin{pmatrix} \beta \\ -\alpha \end{pmatrix}_y \frac{r \sin \theta_k}{\sqrt{\beta_y}}.
\end{aligned}
\end{equation}
The particle trajectories along $2K$ rays in the $(x,y)$ Floquet space starting at the origin are considered. The angles between these rays and the $x$ axes are
\begin{equation}
    \theta_k = k\pi/K \text{ for } k\in\{0,\dots,2K-1\}.
\end{equation}
Since particles get lost on the vacuum chamber walls, it is reasonable to assume that a realistic DA will not exceed the aperture at the reference energy that would exist if all sextupoles and higher-order magnets were turned off. This assumed upper limit is referred to as the \emph{linear aperture}, and the length it spans on the $k$-th ray is denoted by $\bar L(\theta_k)$. 
Similarly, the length that the DA at a relative energy offset $\delta$ spans on this ray is denoted by 
\begin{equation}
    L(\theta_k,\delta).
    \label{eq:L}
\end{equation}
In order not to reward cases with $L(\theta_k,\delta) > \bar L(\theta_k)$, the \emph{line objective} is defined as
\begin{equation}
f_{k,\delta} = \frac{\max\{0,\bar L(\theta_k) - L(\theta_k,\delta)\} }{ \bar L(\theta_k) }. 
\label{eq:line-objective}
\end{equation}
Both $\bar L(\theta_k)$ and $L(\theta_k,\delta)$ are computed using the biased binary search as presented in~\cite{ipac20}. 
For $\bar L_k$ the dimensionless initial length in Floquet space is set to a sufficiently large reference radius of \num{0.01}. The initial radius for $L(\theta_k,\delta)$ is then set to $\bar L_k$.
The \emph{DA objective} for a given relative energy offset $\delta$ is defined as
\begin{equation}
  \mathrm{DA}_\delta = \frac{1}{2K} \sum_{k=0}^{2K-1} f_{k,\delta}^2.\label{eq:DA-delta}
\end{equation} 
In a flat lattice there is vertical symmetry of the aperture area, so \eqref{eq:DA-delta} becomes
\begin{equation}
  \mathrm{DA}_\delta = \frac{1}{2K} \left( f_{0,\delta}^2 + f_{K,\delta}^2 + 2 \sum_{k=1}^{K-1} f_{k,\delta}^2\right).
  \label{eq:DA-delta-2}
\end{equation} 
Due to the normalizations in  Eqs.~\eqref{eq:line-objective} and \eqref{eq:DA-delta}, the DA objective is always in $[0,1]$.

In this paper the on-momentum and off-momentum DA objectives are considered. In particular, 
\begin{equation}
    \mathrm{DA}_{-\delta},\ \mathrm{DA}_{0} \text{ and } \mathrm{DA}_{\delta},
    \label{eq:DA-delta-3}
\end{equation}
where $\delta = 0.03$.

\subsection{Crossing tune resonances\label{sec:crossing-tune-resonances}}

The DA objectives in Eq.~\eqref{eq:DA-delta-3} are computed to ensure a sufficiently large aperture region in phase space. 
Unfortunately, the binary search used to compute the line objectives in Eq.~\eqref{eq:line-objective} 
is sufficient only when the aperture region is simply connected.
This is not always the case, especially when particles cross tune resonances. Therefore, additional requirements that take into account the crossing of tune resonances need to be defined. For this, the tune
\begin{align}
  \vec \nu(x, y, \delta) = \begin{pmatrix} \nu_x(x,y,\delta) \\ \nu_y(x,y,\delta) \end{pmatrix}
\end{align}
is considered as a function of the initial positions $x$, $y$ in transverse Floquet space and the relative energy deviation $\delta$.

\subsubsection{Chromatic tune footprint\label{sec:ctfp}}

For a sufficient energy acceptance it is beneficial to constrain the variation of tunes so that no low-order resonances are crossed. In particular, in this paper the tune footprint is constrained inside the triangle formed by three intersecting 2nd order resonances around the on-momentum tune. In the case of the considered SLS~2.0 lattice, the vertices of this triangle (see Fig.~\ref{fig:footprints} on p.\pageref{fig:footprints}, red lines and part of $x$ axis) are
\begin{equation}
(39,15) - (39.5,15.5) - (39.5,15).
\label{eq:triangle}
\end{equation}
To prevent particles from getting lost on resonance stop-bands, a margin of \num{0.025} around the resonance lines is used in this work (see Fig.~\ref{fig:footprints}, inner black triangle).

The tune footprint is approximately computed by sampling the energy-dependent tunes $\nu_x$ and $\nu_y$ at the energy offsets
\begin{equation}
\delta_p= \frac{p \cdot \delta_\mathrm{max}}P,\hspace{8pt} p = -P, \dots, P.
\end{equation} 
There are cases for which the particle motion is unstable and the tunes cannot be computed. Denoting by $g(\vec \nu)$ the squared Euclidean distance of $\vec \nu$ from the aforementioned triangle (see Eq.~\eqref{eq:triangle}), and taking into account that $\vec \nu(0,0,\delta_0) = \vec \nu(0,0,0)$ is known and that 
\begin{equation}
    g\left(\vec\nu(0,0,0)\right) = 0,
\end{equation}  
the \emph{tune footprint distance} is defined as
\begin{equation}
 \texttt{ctfp} = \sum_{\stackrel{p \neq 0}{\text{ computable}}} g\left(\vec \nu(0,0,\delta_p)\right).
 \label{eq:ctfp}
\end{equation}
Furthermore, as in~\cite{ehrlichman}, 
\begin{equation}
\texttt{unstable}_\pm = 1 - |\delta_{u,\pm}|/\delta_\text{max},
\label{eq:unstable}
\end{equation}
is defined, with $\delta_{u,+}$ and $\delta_{u,-}$ denoting the first (i.e., smallest in magnitude) positive and negative values, respectively, for which the tunes are located outside the triangle or not computable.

The value of $P$ needs to be high enough to achieve a sufficient resolution in tune space, without superfluous computational overhead -- in this paper 
$P = 25$ is used. Furthermore, $\delta_\mathrm{max} = 0.05$ is used.

\subsubsection{Amplitude-dependent tune shifts (ADTS)\label{sec:adts}}

The betatron oscillation is nonlinear and thus anharmonic.
Therefore, a number of different amplitudes have to be considered. 
In this work, to achieve a sufficient resolution,
$Q=20$ equidistant points are taken on each of the following two line segments in the transverse Floquet plane: the horizontal line segment 
(see the sentence containing Eq.~\eqref{eq:L} for the definition of $L(\cdot,0)$)
\begin{equation}
\big\{(t,\Delta)\ |\ t \in\ ]0,L(0,0)]\big\}    
\end{equation}
and the vertical line segment 
\begin{equation}
\big\{(\Delta,t)\ |\ t\in\ ]0,L(\pi/2,0)])\big\}.
\end{equation} 

In these points (with $\delta = 0$) the tunes are computed as the fundamental frequencies of turn-by-turn data in each plane.
To compute these frequencies, the FFT of 128 tracked turns with zero padding to 512 samples is used. To excite both oscillation modes $\Delta=\num{e-4}$ is used, offsetting these line segments from the original rays of the DA computation.

As in the case of computing the tune footprint distance \texttt{ctfp} in Eq.~\eqref{eq:ctfp}, the squared Euclidean distance of the tune in these point from the triangle formed by the 2nd order resonances (see Eq.~\eqref{eq:triangle}) is subsumed into the \emph{amplitude-dependent tune footprint distance}
\begin{align}
 \texttt{adts} &= \sum_{\stackrel{q = 1}{\text{computable}}}^{Q} g\left(\vec \nu(x_q,\Delta,0)\right) +\nonumber\\
 &+ \sum_{\stackrel{q = 1}{\text{computable}}}^{Q} g\left(\vec
 \nu(\Delta,y_q,0)\right).
 \label{eq:adts}
\end{align}
Here `computable' refers to the tracked particle not being lost in 512 turns.

\subsection{Search space\label{sec:search-space}}

Sextupoles are mainly used to compensate chromaticity, but they also limit the on-momentum transverse DA. 
To have the possibility to extend the DA limits, more than two sextupole families are used.
The strengths of these two sextupole families are subsumed into a vector of tuning sextupole strengths $\vec t = (t_1,t_2)$, whose linear relationship with chromaticity is quantified by the matrix $\mathbf T$.
The sextupole strengths of the remaining families are grouped into a vector $\vec \kappa$, and their influence on (linear) chromaticity $\vec \xi = (\xi_x, \xi_y)$ is characterised by a matrix $\mathbf M$, so that
\begin{equation}
  \vec \xi = \mathbf M \vec \kappa + \mathbf T \vec t + \vec \xi_\text{ua},
  \label{eq:tuning}
\end{equation}
where $\vec \xi_\text{ua}$ is the chromaticity of the unaltered lattice. By magnet design, the applicable sextupole strength is limited to some interval $[-\kappa_\text{max},\kappa_\text{max}]$.

In addition to the sextupole strengths $\vec \kappa$, the chromaticity is also taken to be a part of the search space. To prevent head-tail instability, it must be non-negative. On the other hand, the upper limit $\xi_\mathrm{max}$ can be adjusted. 

To sum up, a design point in the search space is
\begin{equation}
  \vec d = \left(\xi_x, \xi_y, \kappa_1,\dots, \kappa_5\right),
\label{eq:design-point}
\end{equation}
where 
\begin{equation}
  \label{eq:dvar-bounds}
  \begin{aligned}
    \xi_{x, y} \in [0,\xi_\mathrm{max}], \hspace{8pt} \kappa_i \in \left[-\kappa_\text{max},\kappa_\text{max}\right].
  \end{aligned}
\end{equation}
The SLS~2.0 sextupoles have a bore of \SI{22}{mm} and a maximum poletip field of \SI{0.71}{T} at \SI{2.7}{GeV}, which corresponds to $\kappa_\text{max} = \SI{650}{m^{-3}}$. Furthermore, in this work $\xi_{\rm max}$ is set to 1.

\subsection{Multiobjective optimization problem}

Summing up the three preceding sections, the constrained multiobjective optimization problem considered in this paper is (see Eqs.~\eqref{eq:DA-delta-2}, \eqref{eq:DA-delta-3}, \eqref{eq:unstable} and \eqref{eq:design-point})
		\begin{equation}
		\underset{\vec{d}}{\text{min}}\ \big(\overbrace{\mathrm{DA}_{-\delta}}^{F_1},\overbrace{\mathrm{DA}_{0}}^{F_2},\overbrace{\mathrm{DA}_{\delta}}^{F_3},\overbrace{\texttt{unstable}_{\mp}}^{F_4,F_5}\big),\label{eq:CMOOP-1}
		\end{equation}
		subject to (see Eqs.~\eqref{eq:ctfp} and \eqref{eq:adts})
		\begin{equation}t_1, t_2 \in [-\kappa_\text{max},\kappa_\text{max}] \text{ and } \texttt{ctfp} + \texttt{adts} = 0.\label{eq:CMOOP-2}
		\end{equation}
The second constraint ensures that the tune footprint distance \texttt{ctfp} and the amplitude-dependent tune footprint distance \texttt{adts} from the triangle formed by 2nd order resonances (see Eq.~\eqref{eq:triangle}) are both zero.

\section{Multiobjective genetic algorithm (MOGA)\label{sec:moga}}

In Eq.~\eqref{eq:CMOOP-1} multiple objectives have to be optimized simultaneously. There are many multiobjective algorithms for this purpose, such as particle swarm optimization~\cite{keeb:95,kara:05,hoss:09,safranek}, ant colony optimization~\cite{domc:96}, simulated annealing~\cite{kigv:83}, artificial immune system~\cite{cati:02}, differential evolution~\cite{husain,wu} or genetic algorithm~\cite{deb:09}. Multiobjective genetic algorithms (MOGA) are probably the most popular and they have already been successfully applied in the field of particle accelerator physics~\cite{neveu,bazarov:05,hofler:13,kranjcevic,Ineichen2013,edelen}, in particular also for the DA optimization~\cite{ehrlichman,li-yang,yang2011,yang2009,gao,li-cheng}.

A design point $\vec{d}_1$ (see Eq.~\eqref{eq:design-point}) \emph{dominates} $\vec{d}_2$ if it is not worse in any of the objectives (see Eq.~\eqref{eq:CMOOP-1}), and it is strictly better in at least one objective. A MOGA allows independent evaluations of solution candidates and is therefore suitable for parallelization.
In this work a massively parallel implementation of a MOGA~\cite{IneichenPhD,Ineichen2013,kranjcevic} is used to find points that are not dominated by any other point, called \emph{Pareto optimal points}.
The basic steps of a MOGA are shown in Algorithm~\ref{alg:MOGA}.

\begin{algorithm}[H]
\caption{Multiobjective genetic algorithm}\label{alg:MOGA}
\begin{algorithmic}[1]
\State random population of individuals, $\vec{d}_i$ for $i = 1, \dots, M$ \label{alg:MOGA-initialize}
\State compute $\vec F(\vec{d}_i)$ for $i = 1, \dots, M$ \label{alg:MOGA-evaluate}
\While {a stopping criterion not reached} \label{alg:MOGA-cycle}
\For {pairs of individuals $\vec{d}_i$, $\vec{d}_{i+1}$} \label{alg:MOGA-crossover1}
\State crossover($\vec{d}_i$, $\vec{d}_{i+1}$), mutate($\vec{d}_i$), mutate($\vec{d}_{i+1}$) \label{alg:MOGA-crossover2}
\EndFor
\State for each new individual $\vec d_{\mathrm{new}}$, compute $\vec F(\vec{d}_\mathrm{new})$ \label{alg:MOGA-evaluate-new}
\State choose $M$ fittest individuals for the next generation \label{alg:MOGA-selector2}
\EndWhile
\end{algorithmic}
\end{algorithm}

In the context of a MOGA a design point is referred to as an \emph{individual}. First, in line~\ref{alg:MOGA-initialize}, $M$ individuals are chosen uniformly at random from intervals in Eq.~\eqref{eq:dvar-bounds}. In line~\ref{alg:MOGA-evaluate} their objective function values (Eq.~\eqref{eq:CMOOP-1}) are computed.  
Then, in lines~\ref{alg:MOGA-cycle}--\ref{alg:MOGA-selector2}, a number of cycles is performed, each resulting in a new generation. In every cycle, new individuals are created using crossover and mutation operators (lines~\ref{alg:MOGA-crossover1}--\ref{alg:MOGA-crossover2}) and their objective function values are computed (line~\ref{alg:MOGA-evaluate-new}). Finally, in line~\ref{alg:MOGA-selector2}, approximately $M$ fittest individuals are chosen to comprise the new generation. The implementation from~\cite{neveu,IneichenPhD} (called \texttt{opt-pilot}) is used, where the algorithm is implemented in C++ and parallelized using MPI such that a new generation is created (line~\ref{alg:MOGA-selector2}) once the objective function values have been computed for $n$ new individuals (line~\ref{alg:MOGA-evaluate-new}). 

In this paper simulated binary crossover and independent bit mutation are used.

\subsection{Particle tracking and lattice configuration\label{sec:tracy}}

The particle tracking code \texttt{tracy}~\cite{tracy} is used to compute the objective function values (Eq.~\eqref{eq:CMOOP-1}) and constraint violations (Eq.~\eqref{eq:CMOOP-2}) for a given design point $\vec d$ (Eq.~\eqref{eq:design-point}).
\texttt{tracy} is a flexible and well-tested beam dynamics library that was also used for SLS~\cite{sls-tracy}.
It uses a 4th-order symplectic integrator~\cite{ruth-forest} for all multipole orders and allows fast tracking of single particles, enabling a trade-off between computation time and accuracy.

For the purpose of this paper the tracking code is modified, including the computation of the ADTS and chromatic tune footprints (see section~\ref{sec:crossing-tune-resonances}) and the DA (see section~\ref{sec:DA}). 
Furthermore, an interface was created so that the values needed in Eqs.~\eqref{eq:CMOOP-1} and \eqref{eq:CMOOP-2} could be obtained by \texttt{opt-pilot}. 

In this paper the current lattice 
for SLS~2.0 is used~\cite{andreas-lattice}. This lattice is based on the multi-bend achromat scheme, including reverse bends and bends with a 3-step longitudinal profile and additional quadrupole focusing. The magnet lattice has a 3-fold symmetry. For on-momentum particles a `virtual' 12-fold symmetry exists due to the proper adjustment of betatron phase advances between sextupoles in the insertion spaces. 
In this work the particles are tracked for 500 turns.

\subsection{Constraint handling\label{sec:constraint-handling}}

Only some randomly chosen individuals (around $\SI{48}{\%}$) satisfy the first constraint in Eq.~\eqref{eq:CMOOP-2}, i.e., their tuning sextupoles are inside of the bounds. In the following, such individuals are called \emph{feasible}.
If an individual is infeasible, its objective function values are not computed. Instead, infeasible individuals are compared based on the severity of their constraint violations. Since the objective function values in Eq.~\eqref{eq:CMOOP-1} are at most one, this can be done by simply setting (see Eq.~\eqref{eq:CMOOP-2})
\begin{align}
  & F_i \leftarrow 2 + \max\left\{0,|t_i| - \kappa_\text{max}\right\} && \text{ for } i=1,2,\\
  & F_i \leftarrow 2 && \text{ for } i = 3,4,5.
\end{align}

On the other hand, for individuals that violate the second constraint in Eq.~\eqref{eq:CMOOP-2} (i.e., when tune resonances are crossed) the objective function value computed in line~\ref{alg:MOGA-evaluate} or \ref{alg:MOGA-evaluate-new} of Algorithm~\ref{alg:MOGA} is penalized as
\begin{equation}
F_i \leftarrow F_i + \texttt{penalty},
\end{equation}
where (see Eqs.~\eqref{eq:ctfp} and \eqref{eq:adts})
\begin{equation}
    \texttt{penalty} = \alpha_1\cdot\texttt{ctfp} + \alpha_2\cdot\texttt{adts}.
\end{equation}
If $\texttt{penalty} \geq 1$ at least one of the tune footprints extends so far outside the triangle that this individual cannot be considered better than all infeasible individuals. Therefore, the possibility that the penalized objectives of this feasible individual are compared with constraint violations of an infeasible individual is allowed. This either results in the standard behavior, i.e., the feasible individual being chosen, or the infeasible individual being chosen -- in which case its tuning sextupoles are likely close to the admissible bounds.

In this work $\alpha_1 = 0.01$ and $\alpha_2 = 1$ are used.

\subsection{Results\label{sec:opt-pilot-results}}

All computations in this paper are run on Intel Xeon Gold 6152 nodes of the PSI Merlin cluster.
An optimization using \texttt{opt-pilot} run for \SI{48}{h} on three nodes (i.e., 132 processes), with $M = 300$ and $n = 130$ (see section~\ref{sec:moga} above~\ref{sec:tracy}), computed 829 generations. The quality of a generation is quantified by counting the number of distinct design points in that generation which satisfy the constraints in Eq.~\eqref{eq:CMOOP-2} and have all of the objective function values better than those of the manually obtained solution. The values for a few representative generations, including the last one, are shown in Table~\ref{Tab:b046_opt-pilot_nof-better}.  
In particular, since the values of the nonnegative objective functions $F_4$ and $F_5$ for the manually obtained solution are zero, they are also zero for all of the newfound points counted in Table~\ref{Tab:b046_opt-pilot_nof-better}. When comparing the results it should also be taken into account that the DAs of the manually obtained solution were optimized beyond the linear aperture limits.

From these 31 good points from the last generation (see Table~\ref{Tab:b046_opt-pilot_nof-better}) three points are chosen based on different criteria and their objective function values are compared to those of the manually obtained solution in Table~\ref{Tab:b046_opt-pilot}.
The design point called \texttt{point-1} is chosen because it has the lowest value of the objective $F_1$, $\textrm{DA}_{-0.03} = 0.021$, and \texttt{point-2} because it has the lowest value of $F_2$ and, coincidentally, also of $F_3$, with values $\textrm{DA}_{-0.03} = 0.001$ and $\textrm{DA}_{0.03} = 0.002$, respectively. The design point \texttt{point-3} was chosen because it improves all three objectives by a comparable amount (for all of these points $F_4 =  F_5 = 0$).

To sum up, using the massively parallel implementation \texttt{opt-pilot} of a MOGA, many design points with very good objective function values are found, but the run time (\SI{48}{h}) is quite long. 

    \begin{table}[h!]
    \centering
    \caption{The number of design points in a specific generation that satisfy the constraints in Eq.~\eqref{eq:CMOOP-2} and have all of the objective function values better than those of the manually obtained solution, referred to as the `design solution'. Generation 829 is the last one that is considered because the optimization was stopped after \SI{48}{h}. All of the objective function values are computed with 500 turns in \texttt{tracy}.}\vspace*{5pt}
    \label{Tab:b046_opt-pilot_nof-better}
    \begin{ruledtabular}
    \begin{tabular}{l|c|c|c|c|c|c}
    generation & $100$ & $200$ & $300$ & $400$ & $500$ & $829$ \\ \hline
    nof better  & $1$ & $10$ & $17$ & $18$ & $26$ & $31$ \\ 
    \end{tabular}
    \end{ruledtabular}
    \end{table}

    \begin{table}[h!]
    \centering
    \caption{A comparison of the manually obtained `design solution' with three good points found in the optimization with \texttt{opt-pilot}. The objective function values are computed with 500 turns in \texttt{tracy}, and all of these design points satisfy the constraints in Eq.~\eqref{eq:CMOOP-2}. Out of the 31 design points in generation 829 computed in \SI{48}{h} (see also Table~\ref{Tab:b046_opt-pilot_nof-better}), \texttt{point-1} is chosen as the design point that has the lowest value of $F_1$ and \texttt{point-2} is chosen to have the lowest value of $F_2$ (coincidentally, it also has the lowest value of $F_3$). \texttt{point-3} is chosen to improve all of these three objectives by a comparable amount. The column labeled `gen' shows the generation in which the specific point was found.}\vspace*{5pt}
    \label{Tab:b046_opt-pilot}
    \begin{ruledtabular}
    \begin{tabular}{l|c|c|c|c|c|c}
    Objective & $F_1$ & $F_2$ & $F_3$ & $F_4$ & $F_5$ & gen \\ \hline
    design solution \hspace{3pt}  & $0.032$ & $0.004$ & $0.011$ & $0$ & $0$ & - \\
    \texttt{point-1}  & \cellcolor{blue!10!white}{$0.021$} & $0.003$ & $0.010$ & $0$ & $0$ & 763 \\ 
    \texttt{point-2}  & $0.031$ & \cellcolor{blue!10!white}{$0.001$} & \cellcolor{blue!10!white}{$0.002$} & $0$ & $0$ & 769 \\ 
    \texttt{point-3}  & $0.025$ & $0.001$ & $0.005$ & $0$ & $0$ & 807\\ 
    \end{tabular}
    \end{ruledtabular}
    \end{table}

\section{Building the surrogate model\label{sec:building-surrogate-model}}

The convergence of the optimization method can be improved by using, e.g., $k$\nolinebreak-means clustering~\cite{li-cheng}, ANN~\cite{wang-2019} or Gaussian process models~\cite{gaussian-process,gaussian-process-appl} to pre-select the points that need to be evaluated. Alternatively, an ANN surrogate model can be trained to approximate the objective function values and then used in the optimization~\cite{edelen}. 
Due to the encouraging results shown in~\cite{edelen}, available tools and promising preliminary computations, the approach in this paper is based on training and using an ANN surrogate model. 

First, a random feasible sample is created and evaluated using \texttt{tracy}. In particular, around \num{7.5e4} design points (see Eq.~\eqref{eq:design-point}) are chosen uniformly at random from the intervals in Eq.~\eqref{eq:dvar-bounds}. Their feasibility (see section~\ref{sec:constraint-handling}) is then checked according to Eq.~\eqref{eq:tuning} and \num{3e4} feasible points are chosen. The run time for this is negligible. These feasible points are evaluated using \texttt{tracy} and divided into training (\SI{70}{\%}), validation (\SI{20}{\%}) and test set (\SI{10}{\%}). This took \SI{9}{h}~\SI{6}{min} 
on five nodes (220 cores).

Second, this random feasible sample is used to train the ANN surrogate model.
In particular, a feed-forward ANN with $N_\text{layers}$ hidden layers is used. The first hidden layer has $N_\text{neurons}$ neurons while the others have $2 N_\text{neurons}$ neurons. The activation function is \texttt{ReLU} and the loss function is the mean squared error. The model is generated in Python using the \texttt{Keras}~\cite{keras} API on top of \texttt{TensorFlow}~\cite{tensorflow}, with some functionality taken from \texttt{MLLIB}~\cite{mllib}.
The \texttt{Talos} framework~\cite{talos} is used to find good hyperparameters
\begin{equation}
    N_\text{layers}\in\{4,5,6\},\ N_\text{neurons}\in\{32,64,128\}
    \label{eq:hyperparameters-1}
\end{equation}
and the batch size for the stochastic gradient descent algorithm Adam~\cite{adam}
\begin{equation}
    N_\text{batch}\in\{128,256\}.
    \label{eq:hyperparameters-2}
\end{equation}
Other parameters for Adam are the default ones~\cite{adam-defaults} from \texttt{Keras}, including the learning rate 0.001. A comparison of the six best combinations is shown in Table~\ref{tab:talos-comparison}. On one core this took \SI{52}{min}.

\begin{table}[h]
\caption{A comparison of hyperparameters (only the six best combinations). The training is stopped if there is no improvement in the validation loss for 100 epochs.\label{tab:talos-comparison}}
\begin{ruledtabular}
\begin{tabular}{r|r|r|r|r|r}
  & epochs & validation loss & $N_\text{layers}$ & $N_\text{neurons}$ & $N_\text{batch}$  \\ \hline
1 & 454 & 0.002487 &  5 & 64 & 128 \\ 
2 & 554 & 0.002494 &  4 & 64 & 128 \\ 
3 & 330 & 0.002504 &  6 & 64 & 128 \\ 
4 & 415 & 0.002521 &  6 & 64 & 256 \\ 
5 & 513 & 0.002532 &  4 & 64 & 256 \\ 
6 & 922 & 0.002550 &  5 & 32 & 128 \\ 
\vdots & \vdots & \vdots & \vdots & \vdots & \vdots \\ 
\end{tabular}
\end{ruledtabular}
\end{table}

\begin{figure*}
    \centering
    \includegraphics[width=\textwidth,clip,trim={1.5cm 0cm 1.5cm 1.5cm}]{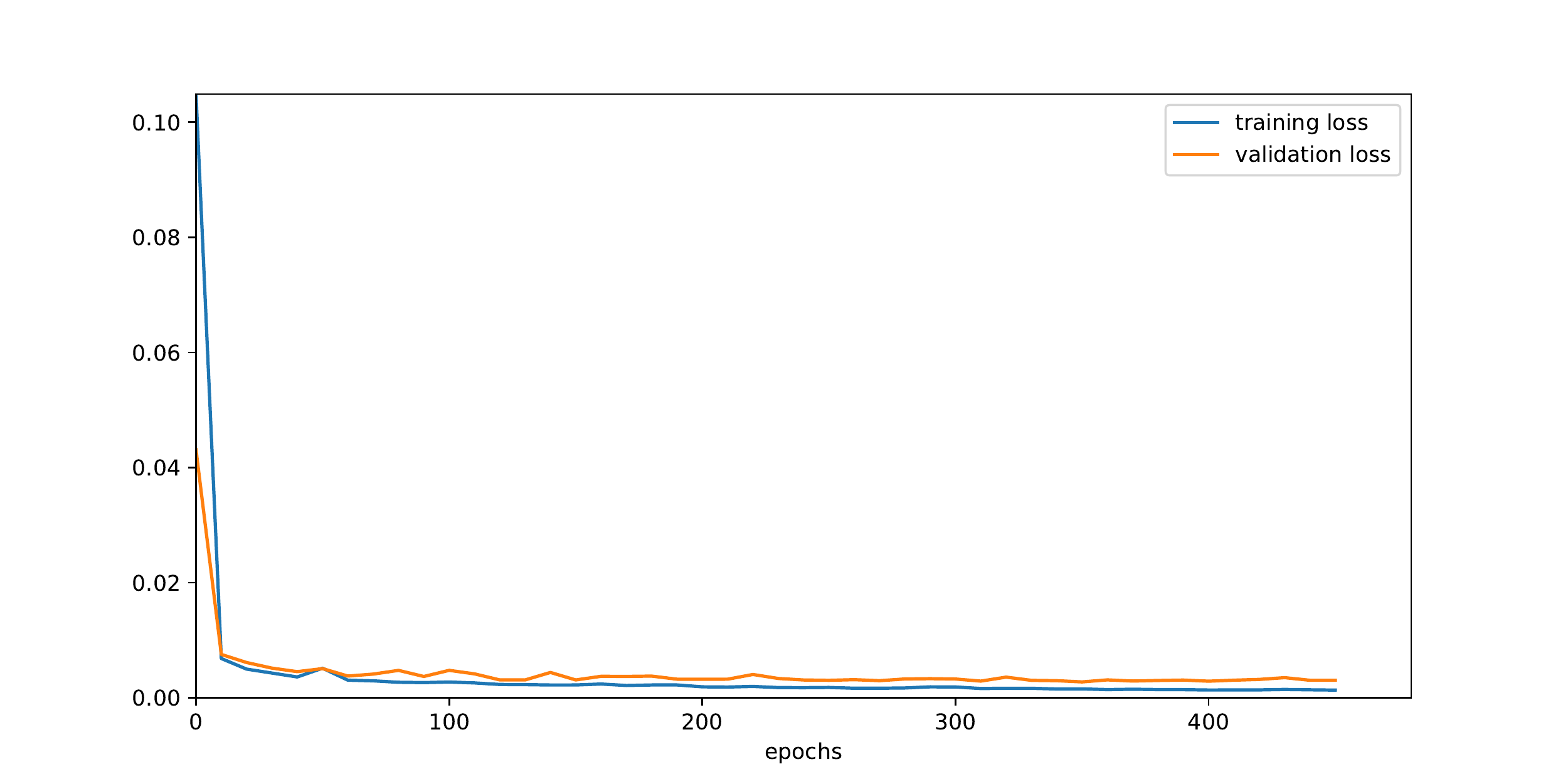}
    \caption{The training loss (blue) and the validation loss (orange) as a function of the number of training epochs. The design point from Eq.~\eqref{eq:design-point} is considered and the functions that are approximated are the ones from Eq.~\eqref{eq:CMOOP-1} and Eq.~\eqref{eq:CMOOP-2}, right (seven in total). The size of the random sample is \num{3e4} and the hyperparameters are: $N_\mathrm{layers} = 5$, $N_\text{neurons} = 64$ and $N_\text{batch} = 128$, with the \texttt{ReLU} activation function.}
    \label{fig:losses}
\end{figure*}

\begin{figure*}
    \includegraphics[width=\textwidth]{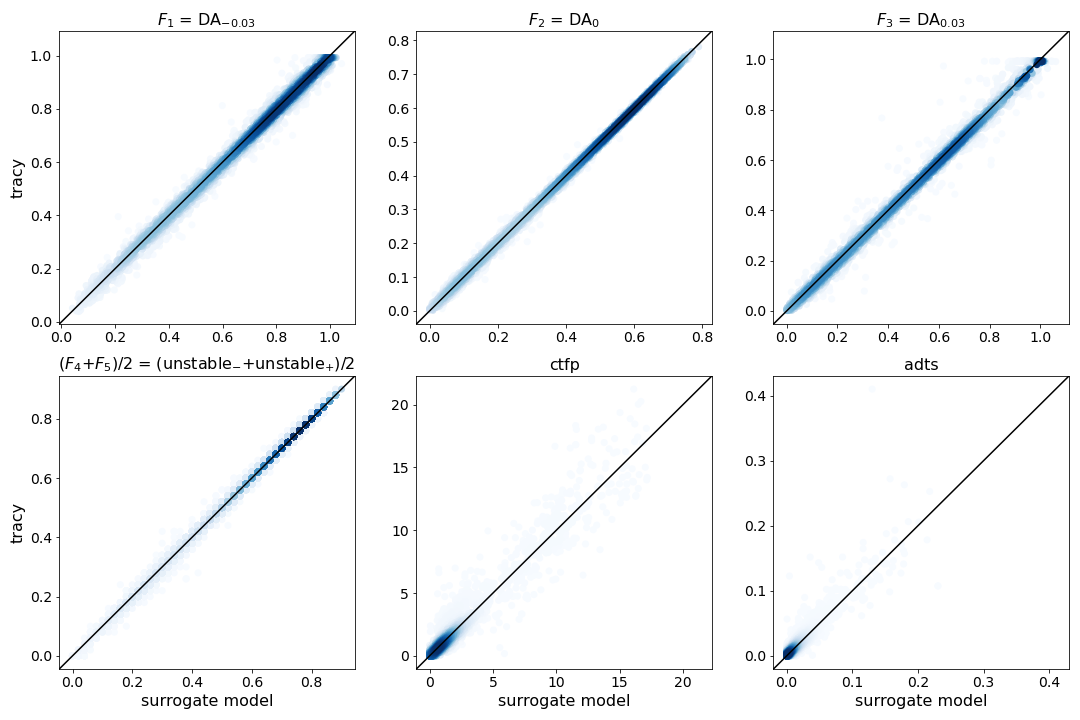}
    \caption{The surrogate model quality on the test set, i.e., random design points that were not used for training. The surrogate model was trained using a training set of size \num{2.1e4} and a validation set of size 6000. Its quality is tested on a test set of size 3000. In each sub-plot the $x$ and $y$ coordinates are the values computed with the surrogate model and \texttt{tracy}, respectively (a point on the line $y = x$ would be perfect agreement). Darker blue colors represent higher design point densities.\label{fig:modelPerformance}}
\end{figure*}

The dependence of the training and validation loss on the number of training epochs for the case with the smallest validation loss (Table~\ref{tab:talos-comparison}), i.e., the hyperparameters 
\begin{equation}
   N_\mathrm{layers} = 5,\ N_\text{neurons} = 64 \text{ and } N_\text{batch} = 128,
\end{equation}
is shown in Fig.~\ref{fig:losses}. 
A comparison of this ANN surrogate model with the particle tracking results in~\texttt{tracy} is shown in Fig.~\ref{fig:modelPerformance}. The comparison is performed on the test set, i.e., random feasible design points that were not used for training.
In each sub-plot the $x$ and $y$ coordinates are the values computed with the ANN surrogate model and \texttt{tracy}, respectively. The line $y = x$ indicates perfect agreement. In the case of $F_4$ and $F_5$ the surrogate model prediction is rounded to the nearest fraction (see Eq.~\eqref{eq:unstable}) and, to facilitate the presentation of the results, the average of these values is shown in the second row, first column. Moreover, for \texttt{ctfp} and \texttt{adts} negative predictions are set to zero. 

\begin{figure*}
    \includegraphics[width=\textwidth]{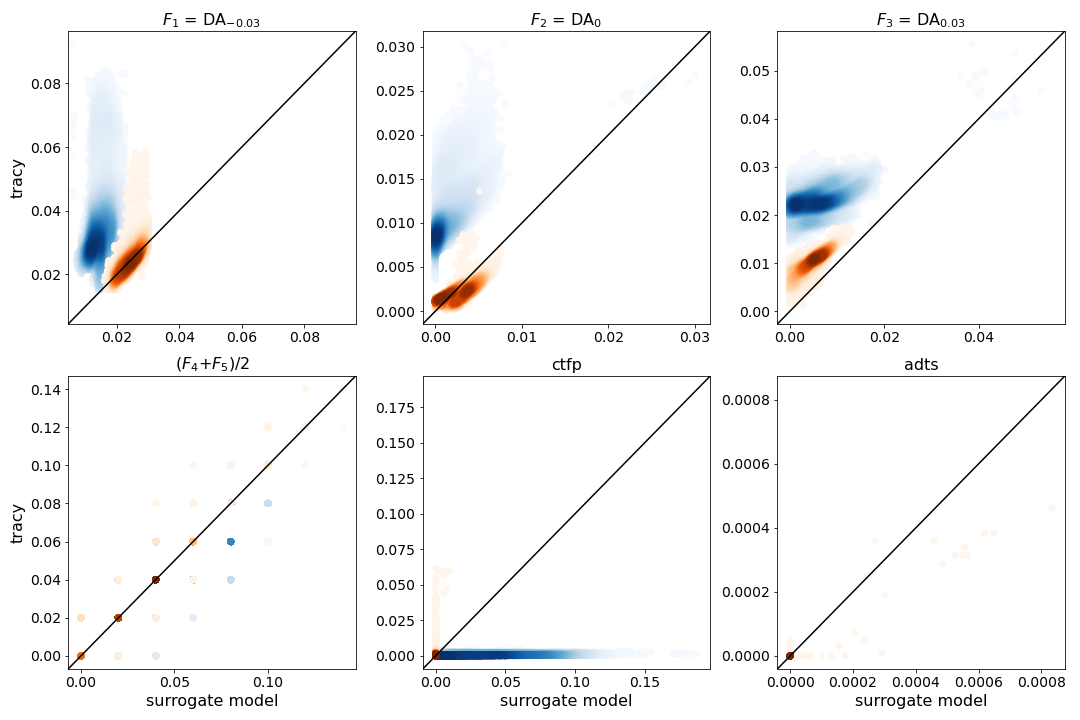}
    \caption{
    Blue color: the quality of the predictions of the surrogate model from Fig.~\ref{fig:modelPerformance} on the design points in generation 1000 (computed with the approach from section~\ref{sec:annsm}). 
    For all points the value of \texttt{adts} is computed as zero using \texttt{tracy} and predicted to be zero using the surrogate model.
    Orange color: the quality of the predictions of the third surrogate model from the approach in section~\ref{sec:annsm-retraining} on the design points in generation 1000. 
    In each sub-plot the $x$ and $y$ coordinates are the values computed with the surrogate model and \texttt{tracy}, respectively (a point on the line $y = x$ would be perfect agreement). Darker colors represent higher design point densities. 
    The benefit of the approach from section~\ref{sec:annsm-retraining} can clearly be seen since: (1) the orange smudges overlap better with the line $y = x$, i.e., the surrogate model predictions at the end of the optimization are more accurate, (2) the orange points have smaller $y$ coordinate values, i.e., better objective function values computed with \texttt{tracy}.\label{fig:gen1000}}
\end{figure*}

\section{Optimizing with the surrogate model\label{sec:optimizing-with-the-surrogate-model}}

In this section, the MOGA implemented in the Python \texttt{pymoo}~\cite{pymoo} module is used for the optimization, with the ANN surrogate model from section~\ref{sec:building-surrogate-model} used to predict the necessary figures of merit (see Eqs.~\eqref{eq:CMOOP-1} and \eqref{eq:CMOOP-2}). 

This is compared to the optimization using \texttt{opt-pilot} as described in section~\ref{sec:opt-pilot-results}, to preserve the solution quality while speeding up the optimization. Additionally, as in section~\ref{sec:opt-pilot-results}, the candidate solutions are again compared with the manually obtained solution.

The crossover, mutation and constraint handling used with \texttt{pymoo} are therefore chosen to be as close as possible to the ones used with \texttt{opt-pilot}.

\subsection{Direct approach\label{sec:annsm}}

As a first approach, an optimization with $M = \num{e4}$ individuals in a generation is run for $1000$ generations. This took \SI{60}{min} on one core. The points in generation $1000$ are then re-evaluated using \texttt{tracy} (\SI{3}{h}~\SI{50}{min} on five nodes) and the comparison is shown in Fig.~\ref{fig:gen1000}, blue color. 
It can be seen from the scale that the objective function values of these design points are very good compared to the values computed for random points (see Fig.~\ref{fig:modelPerformance}). For example, in the case of $F_2 = \textrm{DA}_{0}$ the test sample achieved values up to around $0.8$ and the optimized set of design points always has this value below $0.031$.
However, the agreement between the surrogate model predictions ($x$ axis) and the values obtained with \texttt{tracy} ($y$ axis) is quite poor. In particular, for $F_1$, $F_2$ and $F_3$ the surrogate model prediction is generally much smaller than the value computed in \texttt{tracy} -- the points that seem good during the optimization with the surrogate model turn out to be mediocre.
Therefore, despite the initial surrogate model quality seen in  Fig.~\ref{fig:modelPerformance} as evaluated on random feasible points, the surrogate model quality evaluated on good design points, such as those computed in generation 1000, is not adequate for optimization. For example, none of the design points in generation 1000 have $F_3$ below $0.012$ (Fig.~\ref{fig:gen1000}, first row, third column, blue color). Since the value of $F_3$ for the manually obtained solution is $0.011$ (see Table~\ref{Tab:b046_opt-pilot}), none of these points outperform it.

In total, this approach took around \SI{14}{h}~\SI{48}{min}. For comparison, the optimization with \texttt{opt-pilot} run for \SI{48}{h} on three nodes computed 829 generations (with $M = 300$ and $n = 130$), where 31 points satisfy the constraints and have all objective functions better than the design solution (see section~\ref{sec:opt-pilot-results}). The optimization with the surrogate model is $3.2\times$ faster, but the solution quality is not as good. The comparison is clearly presented in Table~\ref{tab:method-comparison}

To improve the solution quality, the quality of the surrogate model predictions has to be much better for points with good objective function values. To achieve this, in the next section the surrogate model will be re-trained during the optimization. 

\begin{table*}[]
\caption{Solution quality and run time for different optimization methods. `SM' is an abbreviation for `surrogate model'. \texttt{opt-pilot} denotes the massively parallel MOGA implementation combined with \texttt{tracy}, in particular the optimization from section~\ref{sec:opt-pilot-results}. `SM (\num{3e4})' and `SM~$+$~re-train (\num{2e4})' refer to the approaches described in section~\ref{sec:annsm} and \ref{sec:annsm-retraining}, respectively. `SM~$+$~re-train (\num{e4})' and `SM~$+$~re-train (5000)' are described in section~\ref{sec:annsm-retraining-smaller} for $N = 5000$ and $N = 2500$, respectively. The number in the parenthesis is the combined size of the used samples. `nof pts better' refers to the number of design points in the last generation that satisfy the constraints in Eq.~\eqref{eq:CMOOP-2} and have all objectives better than the manually obtained solution.
`re-eval all' refers to the case where all \num{e4} individuals in the last generation are re-evaluated using \texttt{tracy}, which accounts for around \SI{3}{h}~\SI{50}{min} of the total run time. 
`re-eval \SI{10}{\%}' refers to the case where only 1000 of the \num{e4} individuals in the last generation are re-evaluated. The speedup is computed with respect to the \texttt{opt-pilot} approach in the first column.
\label{tab:method-comparison}}
\begin{ruledtabular}
\begin{tabular}{r|c|c|c|c|c}
  & \texttt{opt-pilot} & SM (\num{3e4}) & SM $+$ re-train (\num{2e4}) & SM $+$ re-train (\num{e4}) & SM $+$ re-train ($5000$)  \\ \hline
nof pts better & 31 & 0 & 148 & 368 & 87 \\ 
run time (re-eval all) & \SI{48}{h} & \SI{14}{h}~\SI{48}{min} & \SI{12}{h}~\SI{15}{min} & \SI{8}{h}~\SI{31}{min} & \SI{6}{h}~\SI{33}{min} \\ 
core hours (re-eval all) & 6336 & 2847 & 2325 & 1593 & 1210 \\ 
speedup (re-eval all) & 1.0 & 3.2 & 3.9 & 5.6 & 7.3 \\ 
run time (re-eval \SI{10}{\%}) & - & \SI{11}{h}~\SI{21}{min} & \SI{8}{h}~\SI{52}{min}  & \SI{5}{h}~\SI{5}{min} & \SI{3}{h}~\SI{10}{min} \\ 
core hours (re-eval \SI{10}{\%}) & - & 2089 & 1578 & 838 & 465 \\ 
speedup (re-eval \SI{10}{\%}) & - & 4.2 & 5.4 & 9.4 & 15.1 \\
\end{tabular}
\end{ruledtabular}
\end{table*}

\subsection{Re-training the surrogate model\label{sec:annsm-retraining}}

The second approach, devised to achieve both the run time of the ANN surrogate model optimization and the solution quality of the \texttt{opt-pilot} optimization, is the following: the surrogate model is re-trained during the optimization. To keep the total run time for training the surrogate models low, the re-training is done only two times: first after generation $m_1$ and then after generation $m_2$. The points used for re-training can be chosen as a subset of the random sample used for training the surrogate model in section~\ref{sec:building-surrogate-model} and the points evaluated during the optimization. 
Preliminary computations showed that using only the points in generation $m_i$ ($i\in\{1,2\}$) is not enough to accurately predict the values of new points. On the other hand, using both the initial random sample and some of the points from generation $m_i$ works very well.
To keep the total number of \texttt{tracy} evaluations below \num{3e4}, the surrogate model is trained on \num{e4} random feasible points and re-trained in generation $m_1$ with the random feasible points used previously and 5000 points from generation $m_1$, and again in generation $m_2$ with the \num{1.5e4} points used previously and 5000 additional points from generation $m_2$.
Preliminary computations showed that $m_1 = 50$ and $m_1 = 500$ can be used, due to a more rapid change of the objective function values in the beginning of the optimization. For example, in generation $50$ all of the points had $F_2 = \textrm{DA}_{0} < 0.07$, and the values achieved in 1000 generations are generally comparable to the value of the manually obtained solution, $F_2 = 0.004$ (Table~\ref{Tab:b046_opt-pilot}).

The quality of the predictions of the third surrogate model on the \num{e4} design points from generation 1000 is shown in Fig.~\ref{fig:gen1000}, orange color. 
The total run time, including re-evaluating the entire last generation using \texttt{tracy}, is now around \SI{12}{h}~\SI{15}{min}, which is a speedup of $3.9\times$ compared to the approach from section~\ref{sec:opt-pilot-results}. However, not all points in the last generation need to be re-evaluated -- if only 1000 of these points (i.e., \SI{10}{\%}) are re-evaluated, the total run time is \SI{8}{h}~\SI{52}{min}, which is a speedup of $5.4\times$ (see Table~\ref{tab:method-comparison}). The 1000 points to be re-evaluated can be chosen based on the values of the predictions. 

There are 148 design points in the last generation that satisfy the constraints in Eq.~\eqref{eq:CMOOP-2} and have all of the objective function values better than those of the manually obtained solution, which is significantly more than the $31$ points found in section~\ref{sec:opt-pilot-results} (see Tables~\ref{Tab:b046_opt-pilot_nof-better} and \ref{tab:method-comparison}).

As in Table~\ref{Tab:b046_opt-pilot}, out of these 148 design points, the ones with the lowest value of $F_1$, $F_2$ and $F_3$ are shown in Table~\ref{Tab:notable-points} and referred to as \texttt{point-4}, \texttt{point-5} and \texttt{point-6}, respectively.
The quality of these points is clearly comparable with \texttt{point-1} and \texttt{point-2} from Table~\ref{Tab:b046_opt-pilot}. For example, the lowest value of $F_1 = \textrm{DA}_{-0.03} = 0.020$ is achieved with the new approach for \texttt{point-4}, while \texttt{point-1} has $F_1 = 0.021$. On the other hand, \texttt{point-2} has $F_3 = \textrm{DA}_{0.03} = 0.002$, while the lowest value achieved with the new approach is $F_3 = 0.004$ for \texttt{point-6}. With both approaches the lowest value of $F_2 = \textrm{DA}_{0} = 0.001$ is achieved.
Furthermore, there are design points whose quality is comparable to that of \texttt{point-3} in Table~\ref{Tab:b046_opt-pilot}, e.g., a design point with 
\begin{equation}
 (F_1,F_2,F_3,F_4,F_5) = (0.026, 0.001, 0.004, 0, 0),
 \label{eq:point-12}
\end{equation}
$\texttt{ctfp} = 0$ and $\texttt{adts} = 0$.

    \begin{table}[h!]
    \centering
    \caption{
    The points with the smallest value of $F_1$ (\texttt{point-4,7,10}), $F_2$ (\texttt{point-5,8,11}) and $F_3$ (\texttt{point-6,9,11}), out of all the points (first row in Table~\ref{tab:method-comparison}) that satisfy the constraints in Eq.~\eqref{eq:CMOOP-2} and have all objectives better than the manually obtained solution (first row in Table~\ref{Tab:b046_opt-pilot}).
    All the digits of the objective function values computed with 500 turns in \texttt{tracy} are used for the comparison, and the shaded numbers denote the minimal value of the respective objective before rounding.
    The last column denotes the size of the initial sample. The case with $N = \num{e4}$ is described in section~\ref{sec:annsm-retraining}. The cases with $N = 5000$ and $N = 2500$ are described in section~\ref{sec:annsm-retraining-smaller}.
    }\vspace*{5pt}
    \label{Tab:notable-points}
    \begin{ruledtabular}
    \begin{tabular}{l|c|c|c|c|c|c}
    Objective & $F_1$ & $F_2$ & $F_3$ & $F_4$ & $F_5$ & $N$ \\ \hline
    \texttt{point-4}  & \cellcolor{blue!10!white}{$0.020$} & $0.004$ & $0.011$ & $0$ & $0$ & \num{e4} \\
    \texttt{point-5}  & $0.029$ & \cellcolor{blue!10!white}{$0.001$} & $0.008$ & $0$ & $0$ & \num{e4} \\
    \texttt{point-6}  & $0.028$ & $0.002$ & \cellcolor{blue!10!white}{$0.004$} & $0$ & $0$ & \num{e4} \\
    \texttt{point-7}  & \cellcolor{blue!10!white}{$0.024$} & $0.003$ & $0.008$ & $0$ & $0$ & 5000 \\
    \texttt{point-8}  & $0.030$ & \cellcolor{blue!10!white}{$0.002$} & $0.006$ & $0$ & $0$ & 5000 \\
    \texttt{point-9}  & $0.029$ & $0.002$ & \cellcolor{blue!10!white}{$0.004$} & $0$ & $0$ & 5000 \\
    \texttt{point-10}  & \cellcolor{blue!10!white}{$0.025$} & $0.003$ & $0.011$ & $0$ & $0$ & 2500 \\ 
    \texttt{point-11}  & $0.032$ & \cellcolor{blue!10!white}{$0.001$} & \cellcolor{blue!10!white}{$0.005$} & $0$ & $0$ & 2500 \\
    \end{tabular}
    \end{ruledtabular}
    \end{table}

\subsection{Using a smaller sample for training\label{sec:annsm-retraining-smaller}}

In section~\ref{sec:annsm} a large sample of \num{3e4} random feasible design points was used to illustrate that, regardless of the surrogate model quality on random points, the quality on points with good objective functions is very poor. In section~\ref{sec:annsm-retraining} the combined size of the samples was \num{2e4} instead of \num{3e4}. In this section the size of the samples is further reduced.

The approach is the same as the one from section~\ref{sec:annsm-retraining}.
An initial sample of size $N$ is used to train the first surrogate model. The second surrogate model is trained in generation $m_1 = 50$ using these $N$ points together with $N/2$ of the points found in generation $m_1$. The third surrogate model is trained in generation $m_2 = 500$ using also $N/2$ of the points found in generation $m_2$. While in section~\ref{sec:annsm-retraining} $N=\num{e4}$ was used, this is now lowered to $N = 5000$ and $N = 2500$.

As shown in Table~\ref{tab:method-comparison}, the number of points that satisfy the constraints in Eq.~\eqref{eq:CMOOP-2} and have all of the objective function values better than those of the manually obtained solution is now 368 for $N = 5000$ and 87 for $N = 2500$. The run time including the re-evaluation of the entire last generation is \SI{8}{h}~\SI{31}{min} (speedup $5.6\times$) and \SI{6}{h}~\SI{33}{min} (speedup $7.3\times$), respectively. If only 1000 of the points in the last generation (i.e., \SI{10}{\%}) are re-evaluated using \texttt{tracy}, the speedups in the cases $N = 5000$ and $N = 2500$ are $9.4\times$ and $15.1\times$, respectively. A detailed comparison is shown in Table~\ref{tab:method-comparison}.

Furthermore, in addition to counting `nof pts better' (Table~\ref{tab:method-comparison}), out of these design points the points with the lowest values of $F_1$, $F_2$ and $F_3$ are shown in Table~\ref{Tab:notable-points}. 
Both approaches from this section found numerous points with very good objective function values in a significantly shorter time -- most notably the \SI{3}{h}~\SI{10}{min} for the fastest case (instead of \SI{48}{h} needed for \texttt{opt-pilot}).

\section{Candidate solutions\label{sec:results}}

In this section some of the candidate solutions obtained in sections~\ref{sec:moga} and \ref{sec:optimizing-with-the-surrogate-model} are compared and further analyzed.
In particular, out of the points found in section~\ref{sec:opt-pilot-results} with the massively parallel implementation of a MOGA, \texttt{point-3} (see Table~\ref{Tab:b046_opt-pilot}) is chosen. Out of the points found with the new method in section~\ref{sec:annsm-retraining} ($N = \num{e4}$), the point shown in Eq.~\eqref{eq:point-12} is chosen. Out of the points found with the new method in section~\ref{sec:annsm-retraining-smaller}, using the smallest considered combined sample size ($N = 2500$), \texttt{point-11} (see Table~\ref{Tab:notable-points}) is chosen. All of these points are compared with the manually obtained solution, referred to as the `design solution'.

\begin{figure*}
    \subfloat[Design solution.\label{fig:DA-point0}]{
    \includegraphics[width=0.49\textwidth,clip,trim={1.8cm 1cm 1.3cm 1.3cm}]{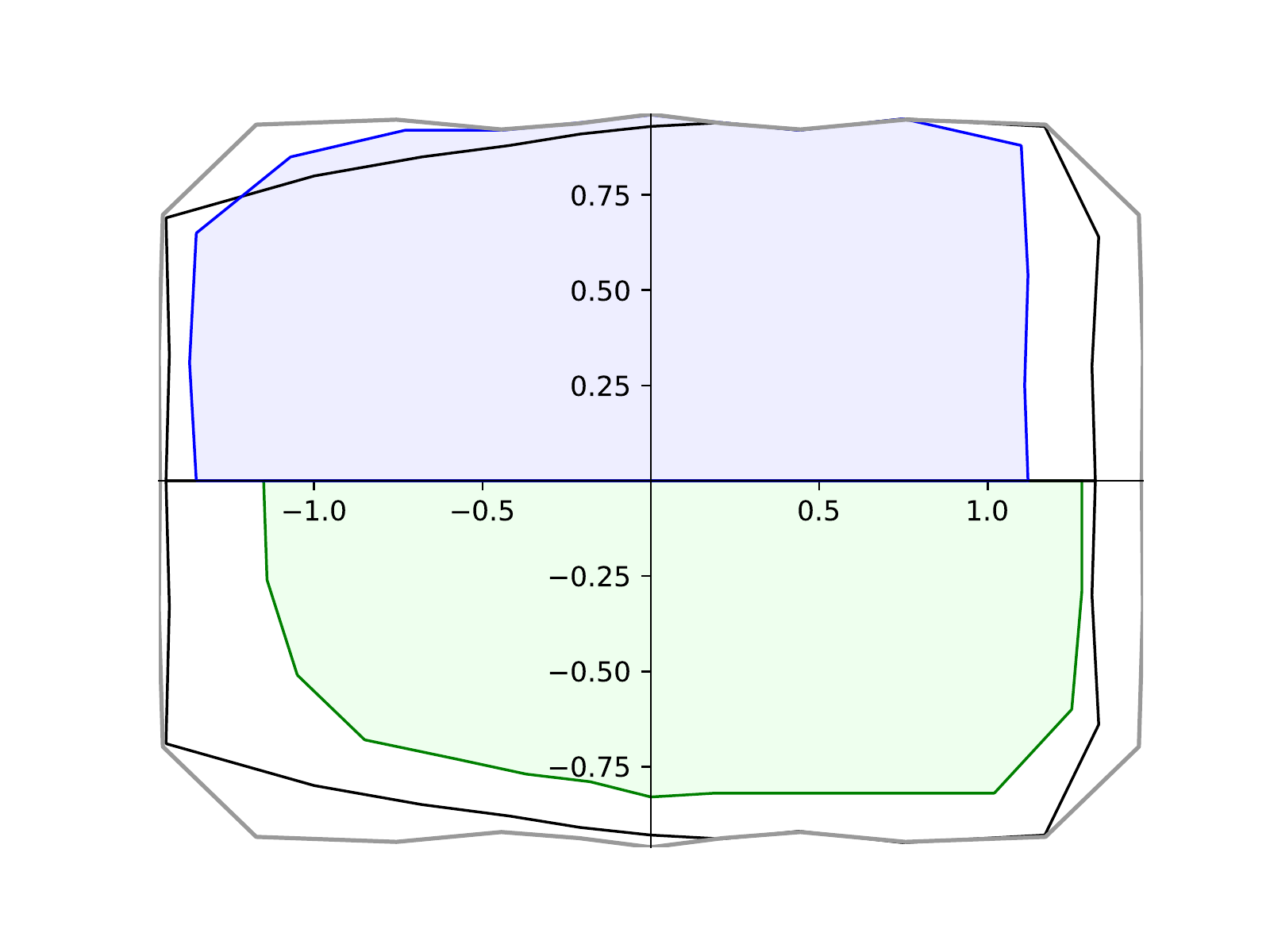}}
    \subfloat[\texttt{point-3} from Table~\ref{Tab:b046_opt-pilot}.\label{fig:DA-point3}]{
    \includegraphics[width=0.49\textwidth,clip,trim={1.8cm 1cm 1.3cm 1.3cm}]{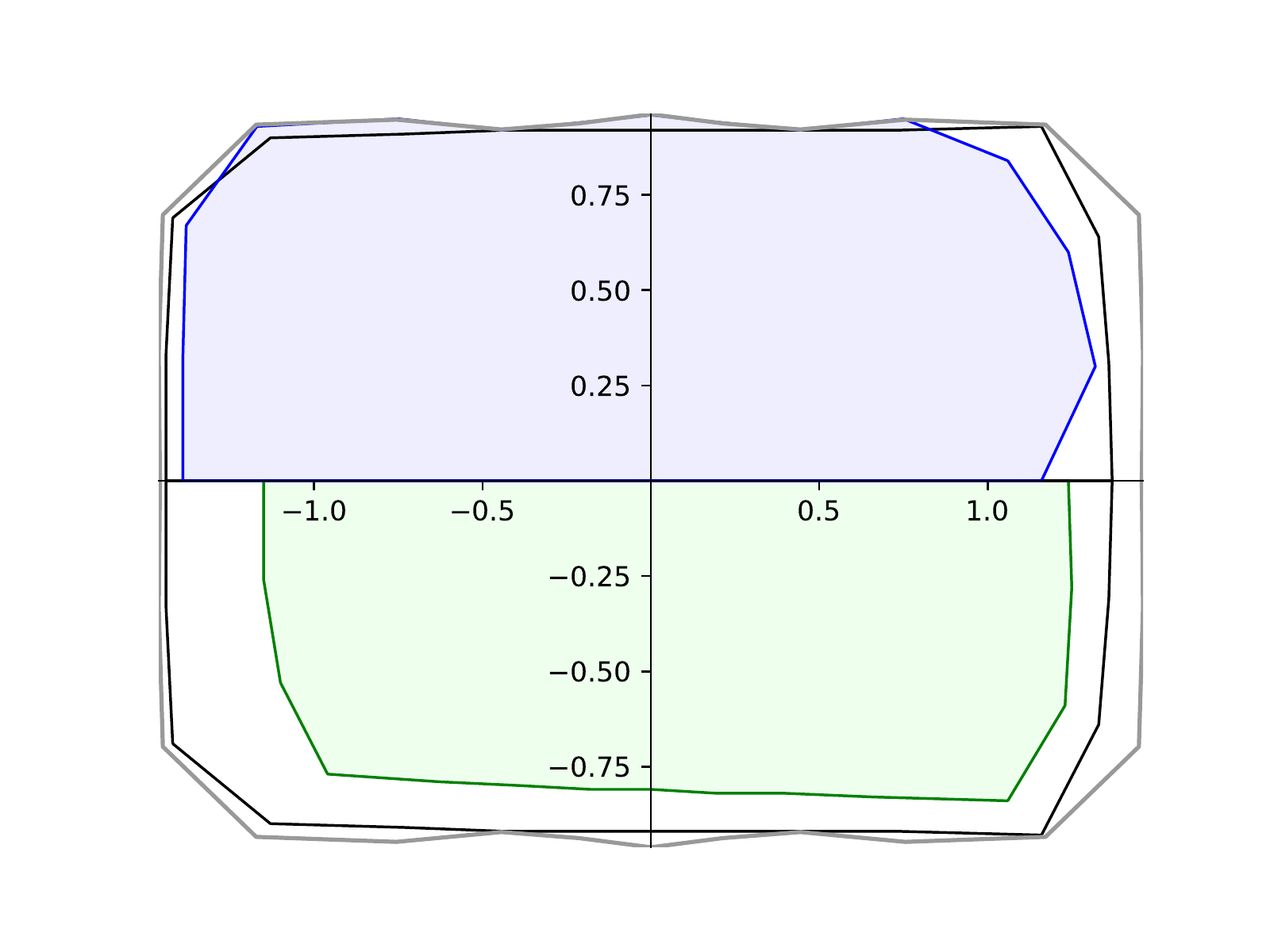}}\\
    \subfloat[The point from Eq.~\eqref{eq:point-12}.\label{fig:DA-point12}]{
    \includegraphics[width=0.49\textwidth,clip,trim={1.8cm 1cm 1.3cm 1.3cm}]{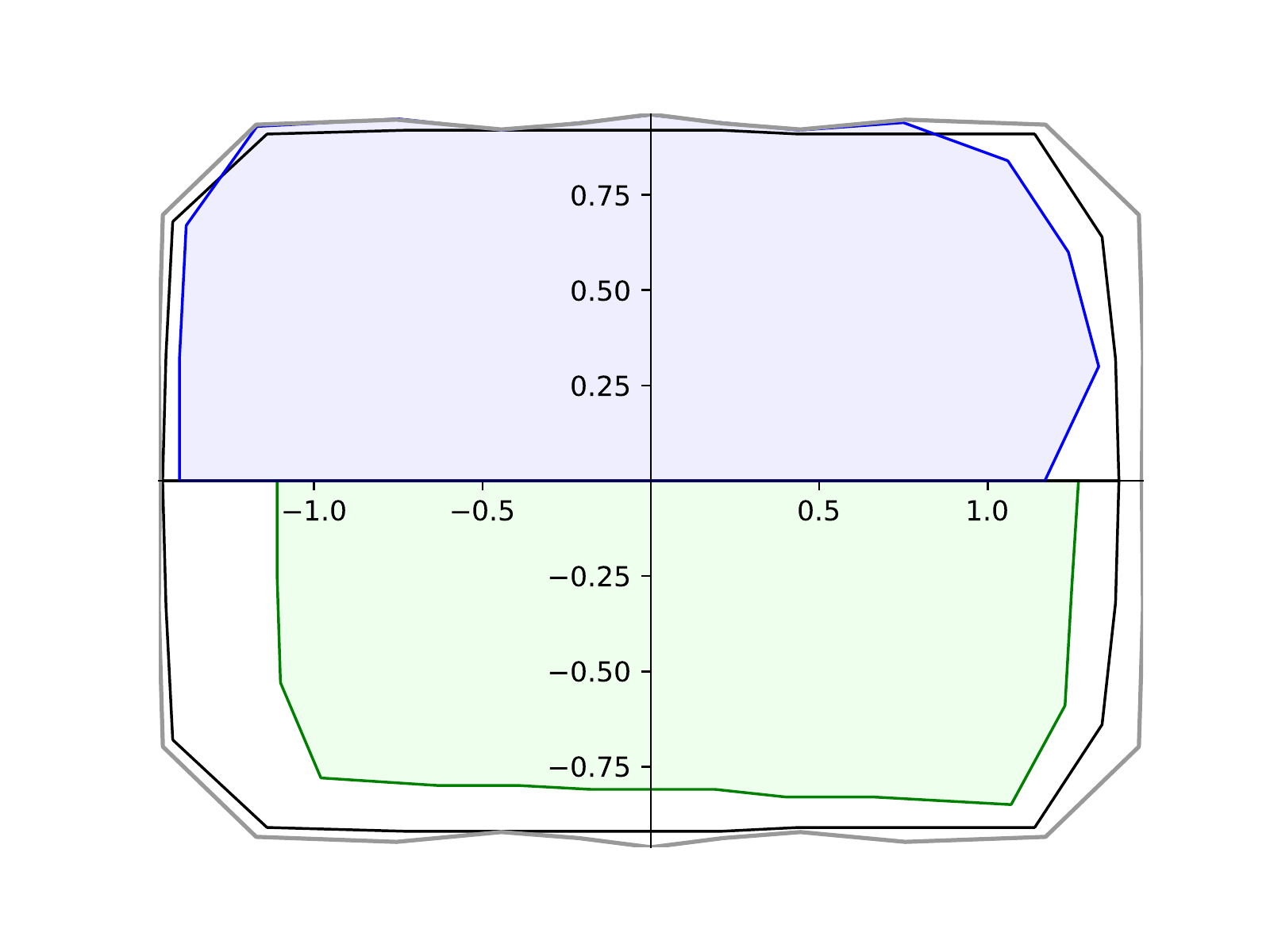}}
    \subfloat[\texttt{point-11} from Table~\ref{Tab:notable-points}.\label{fig:DA-point11}]{
    \includegraphics[width=0.49\textwidth,clip,trim={1.8cm 1cm 1.3cm 1.3cm}]{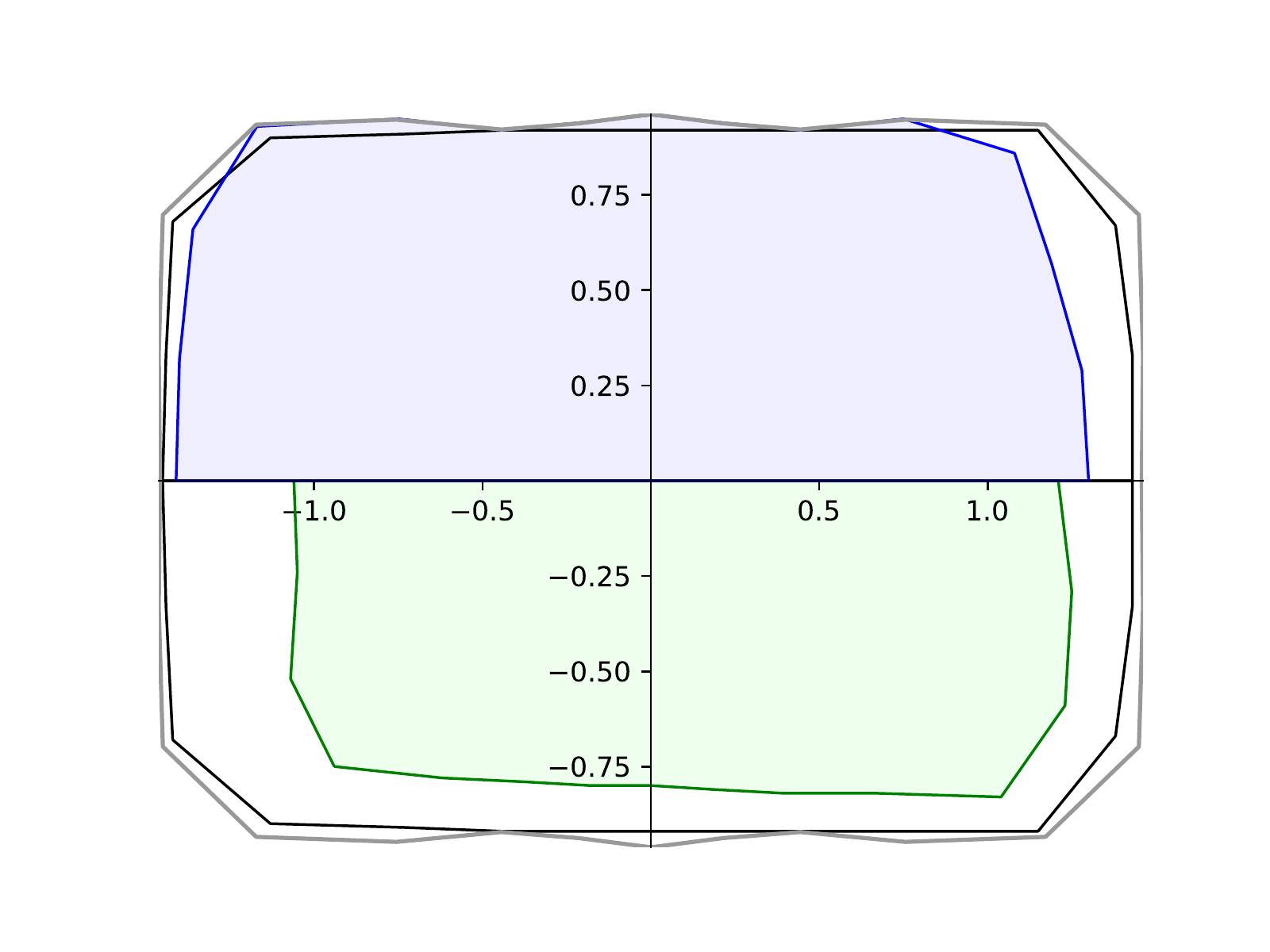}}\\
    \caption{
    Transverse DAs in Floquet space of the solution candidates for $\delta  = -0.03$ (green), $\delta = 0.03$ (blue) and $\delta = 0$ (bold black line), computed using \texttt{tracy} as described in section~\ref{sec:DA}. The particles are tracked for 500 turns.
    For a clearer presentation of the results, only half of the off-momentum apertures is shown. This is sufficient due to machine-plane symmetry.
    Each sub-plot corresponds to one candidate solution:
    the manually obtained solution (referred to as the `design solution', top left), \texttt{point-3} from Table~\ref{Tab:b046_opt-pilot} (top right), the design point from Eq.~\eqref{eq:point-12} (bottom left) and \texttt{point-11} from Table~\ref{Tab:notable-points} (bottom right).
    The relative relationships of the DA area sizes are consistent with the computed objective function values.
    \label{fig:DAs}}
\end{figure*}

For each of these solution candidates, the transverse DAs at three different energies ($\delta \in \{-0.03,0,0.03\}$) are shown in Fig.~\ref{fig:DAs}. In each sub-plot the bold black line shows the boundary of the on-momentum DA, computed with 500 turns in \texttt{tracy} as described in section~\ref{sec:DA}. As indicated by the smaller values of the objective function $F_2 = \textrm{DA}_{0}$, all three candidate solutions have a larger on-momentum DA than the design solution. The green and blue areas show the off-momentum DA for $\delta = -0.03$ and $\delta = 0.03$, respectively. These transverse DAs correspond to the objective functions $F_1 = \textrm{DA}_{-0.03}$ and $F_3 = \textrm{DA}_{0.03}$, respectively. In the case of $\delta = 0.03$ it can clearly be seen that the computed transverse DA for all three new candidate solutions is larger than that of the design solution, as indicated by the smaller values of $F_3$. 
In the case of $\delta = -0.03$ the area of the transverse DA for \texttt{point-11} is of a similar size to that of the design solution. This agrees with the fact that for both of these design points $F_1 = 0.032$ (see Tables~\ref{Tab:b046_opt-pilot} and \ref{Tab:notable-points}). The other two new candidate solutions have lower values of $F_1$ ($F_1 = 0.025$ in Table~\ref{Tab:b046_opt-pilot} and $F_1 = 0.026$ in Eq.~\eqref{eq:point-12}) and the computed areas of the transverse DA at $\delta = -0.03$ for these two design points are larger than that of the design solution, which is again the desired behaviour.

\begin{figure*}
    \centering
    \begin{tabular}{r|ccc}
& DA, $\delta=-0.03$ & DA, $\delta=0$ & DA, $\delta=0.03$ \\ \hline & & & \\[-.8em]
\rotatebox{90}{\quad\quad\quad design solution}&
\includegraphics[width=.63\columnwidth]{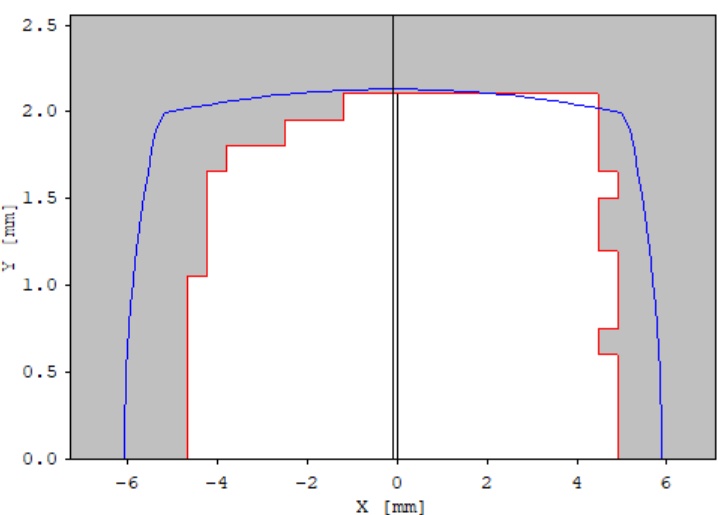}&
\includegraphics[width=.63\columnwidth]{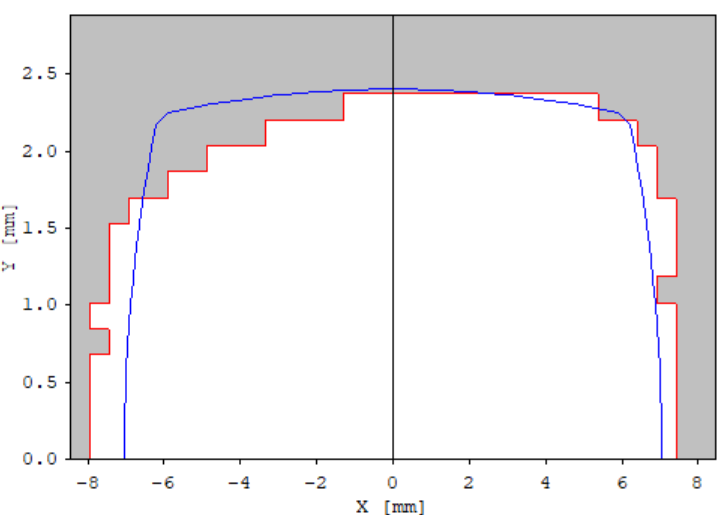}&
\includegraphics[width=.63\columnwidth]{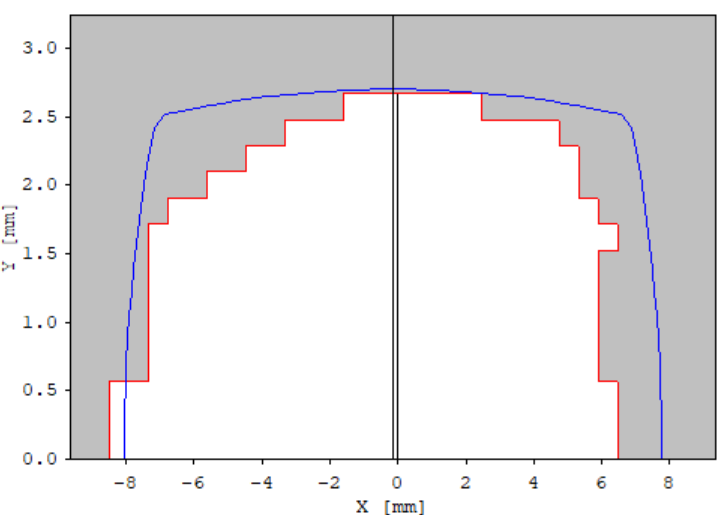}\\
\rotatebox{90}{\quad\quad\quad\quad \texttt{point-3}} &
\includegraphics[width=.63\columnwidth]{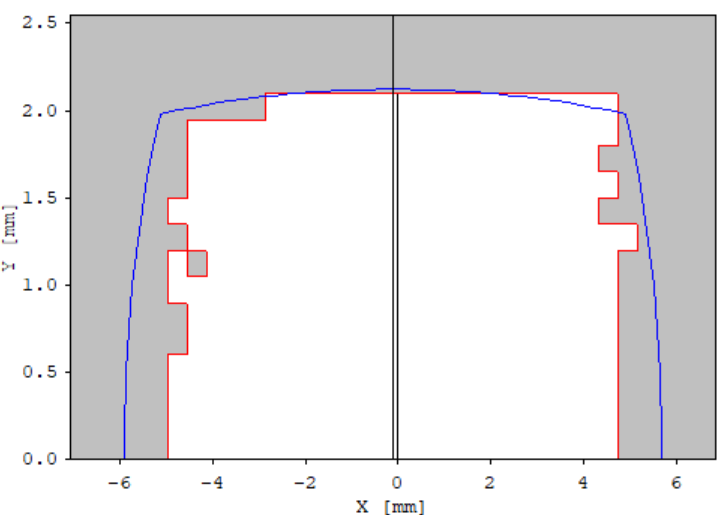}&
\includegraphics[width=.63\columnwidth]{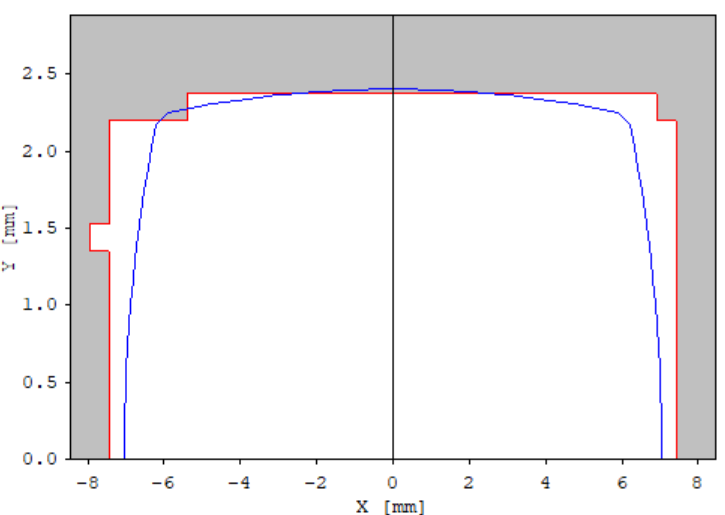}&
\includegraphics[width=.63\columnwidth]{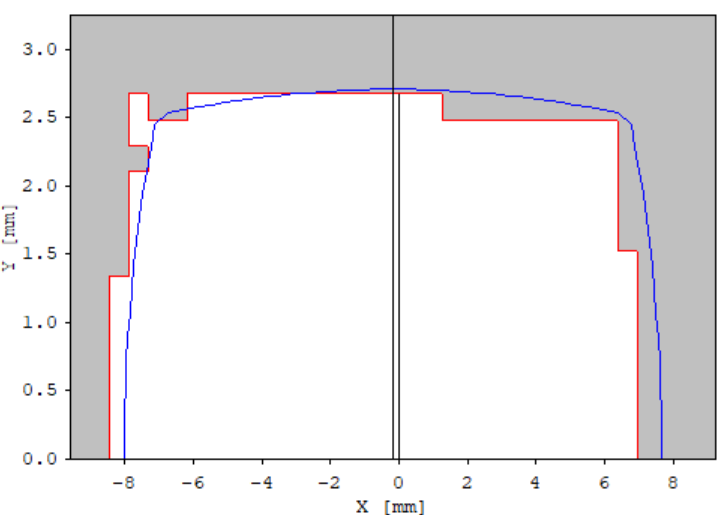}\\
\rotatebox{90}{\quad\quad point from Eq.~\eqref{eq:point-12}} &
\includegraphics[width=.63\columnwidth]{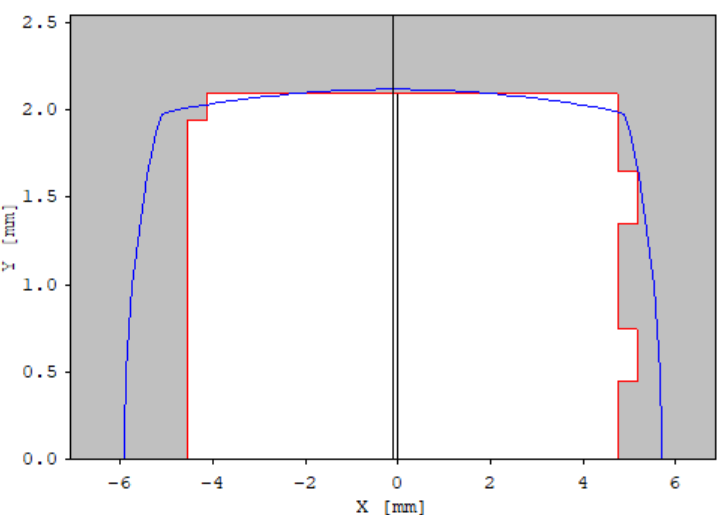}&
\includegraphics[width=.63\columnwidth]{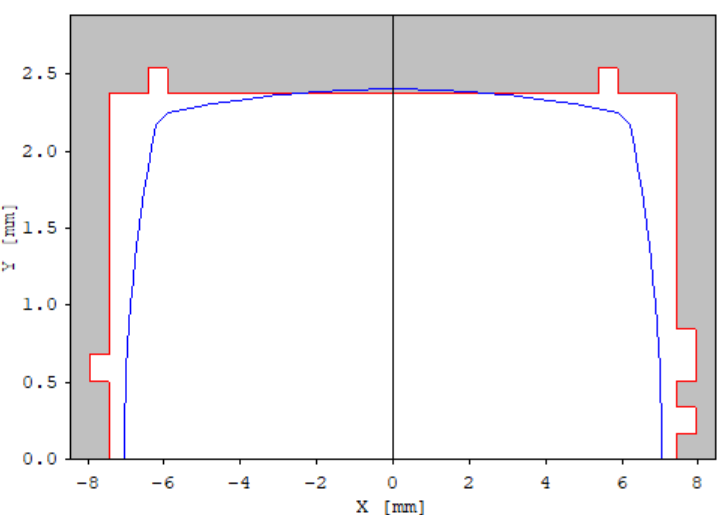}&
\includegraphics[width=.63\columnwidth]{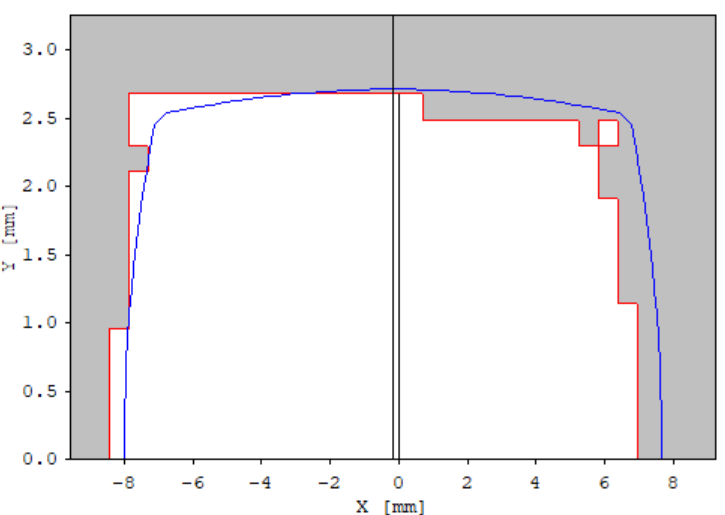}\\
\rotatebox{90}{\quad\quad\quad\quad \texttt{point-11}} &
\includegraphics[width=.63\columnwidth]{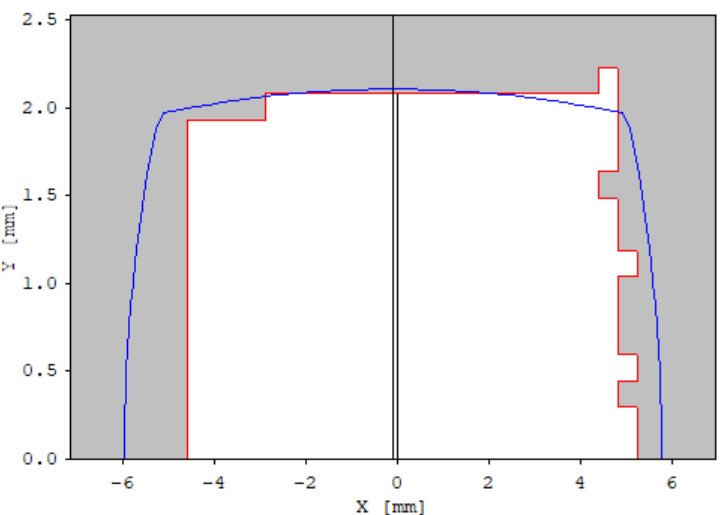}&
\includegraphics[width=.63\columnwidth]{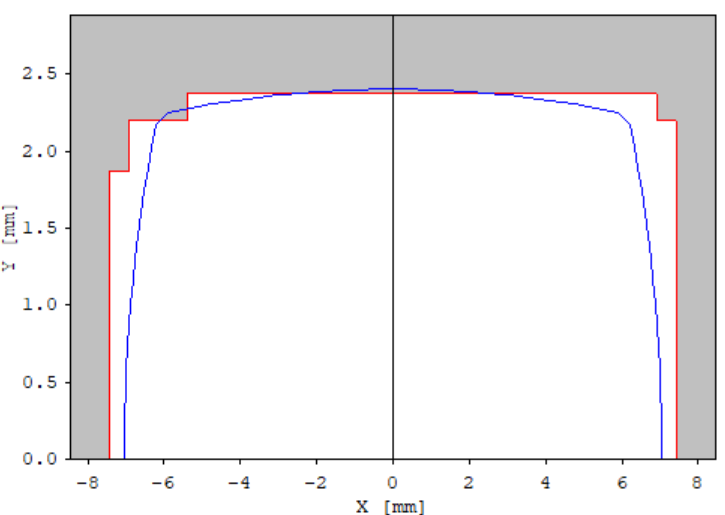}&
\includegraphics[width=.63\columnwidth]{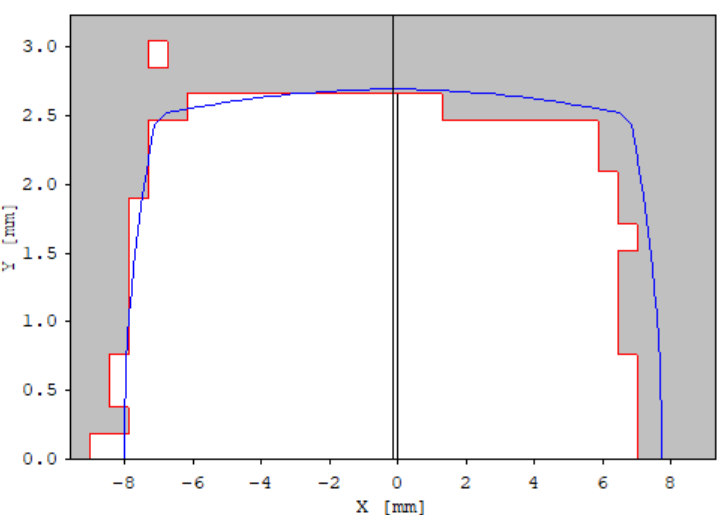}
    \end{tabular}
    \caption{The columns show the transverse DAs at $\delta  = -0.03$, $\delta  = 0$ and $\delta  = 0.03$, recomputed in OPA for: the manually obtained solution (referred to as the `design solution', 1st row), \texttt{point-3} from Table~\ref{Tab:b046_opt-pilot} (2nd row), the design point from Eq.~\eqref{eq:point-12} (3rd row) and \texttt{point-11} from Table~\ref{Tab:notable-points} (4th row). 
    The particles are tracked for 500 turns on a grid in the transverse plane. The results are consistent with the \texttt{tracy} computation in Fig.~\ref{fig:DAs}.
    }
    \label{Tab:dynap_OPA}
    \setlength{\tabcolsep}{3pt}
\end{figure*}

Additionally, the transverse DAs at the three considered energies are computed with OPA~\cite{opa} (Fig.~\ref{Tab:dynap_OPA}). In OPA, for each energy, the transverse DA is sampled on a two-dimensional grid of points.
This results in a better approximation of the DA (cf.~section~\ref{sec:DA}), at the expense of computation time. 
For the considered candidate solutions, the OPA-computed transverse DAs are larger than that of the manually obtained solution. This is in agreement with the relative relationships of the \texttt{tracy}-computed transverse DAs (Fig.~\ref{fig:DAs}).

\begin{figure*}
    \subfloat[Design solution.\label{fig:footprint-point0}]{
    \includegraphics[width=0.245\textwidth]{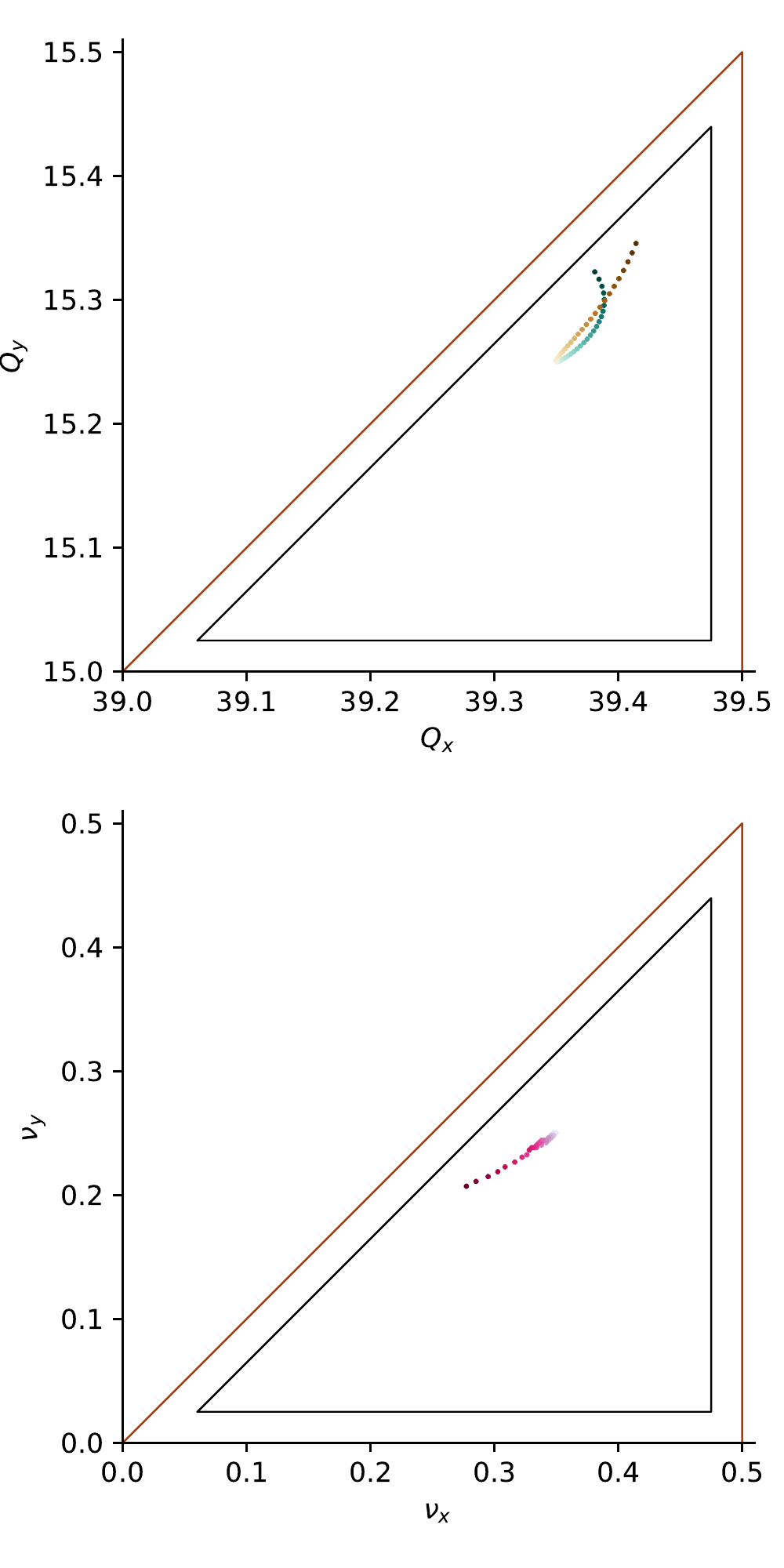}}
    \subfloat[\texttt{point-3} from Table~\ref{Tab:b046_opt-pilot}.\label{fig:footprint-point3}]{
    \includegraphics[width=0.245\textwidth]{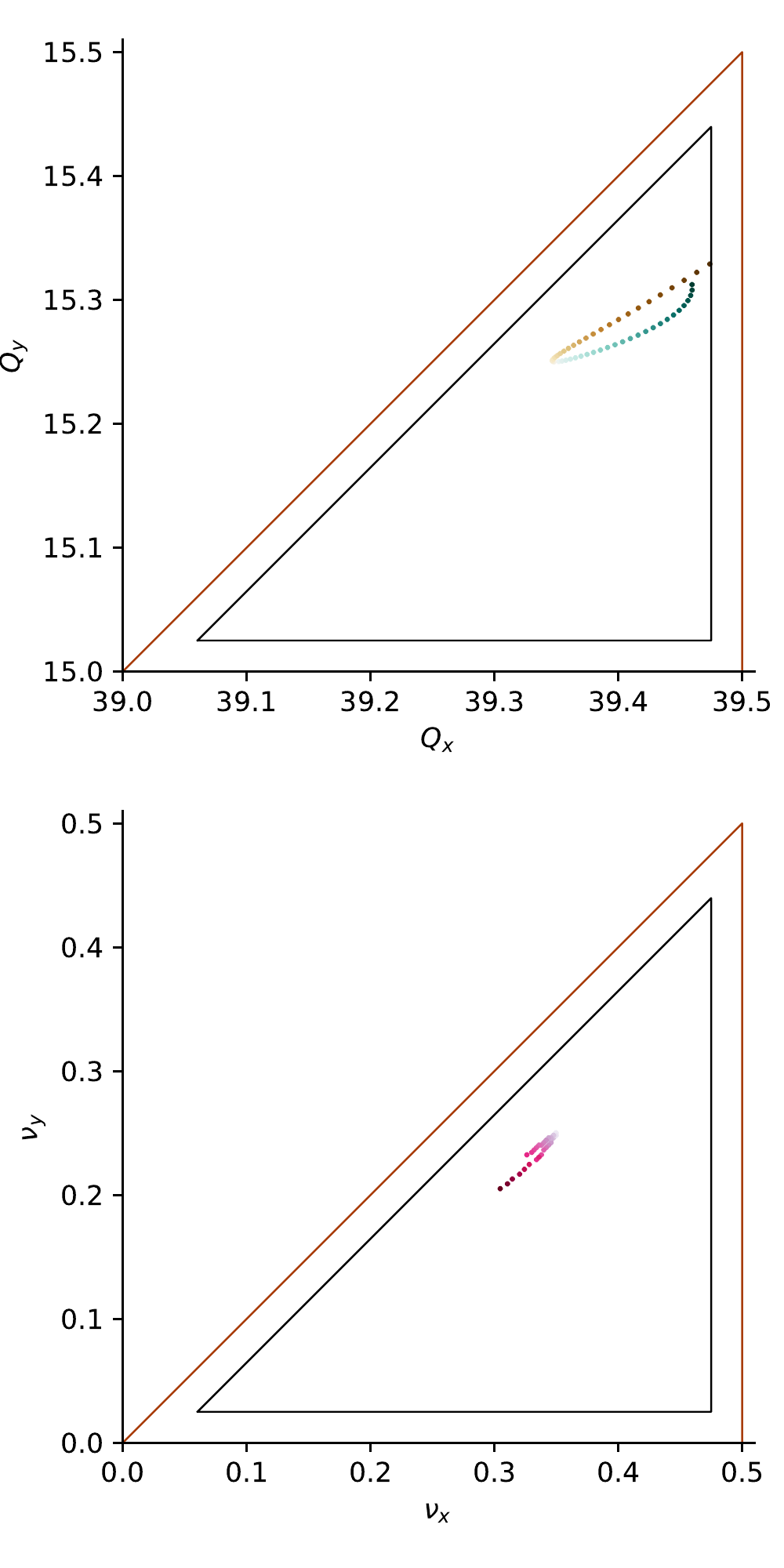}}
    \subfloat[The point from Eq.~\eqref{eq:point-12}.\label{fig:DA-point12}]{
    \includegraphics[width=0.245\textwidth]{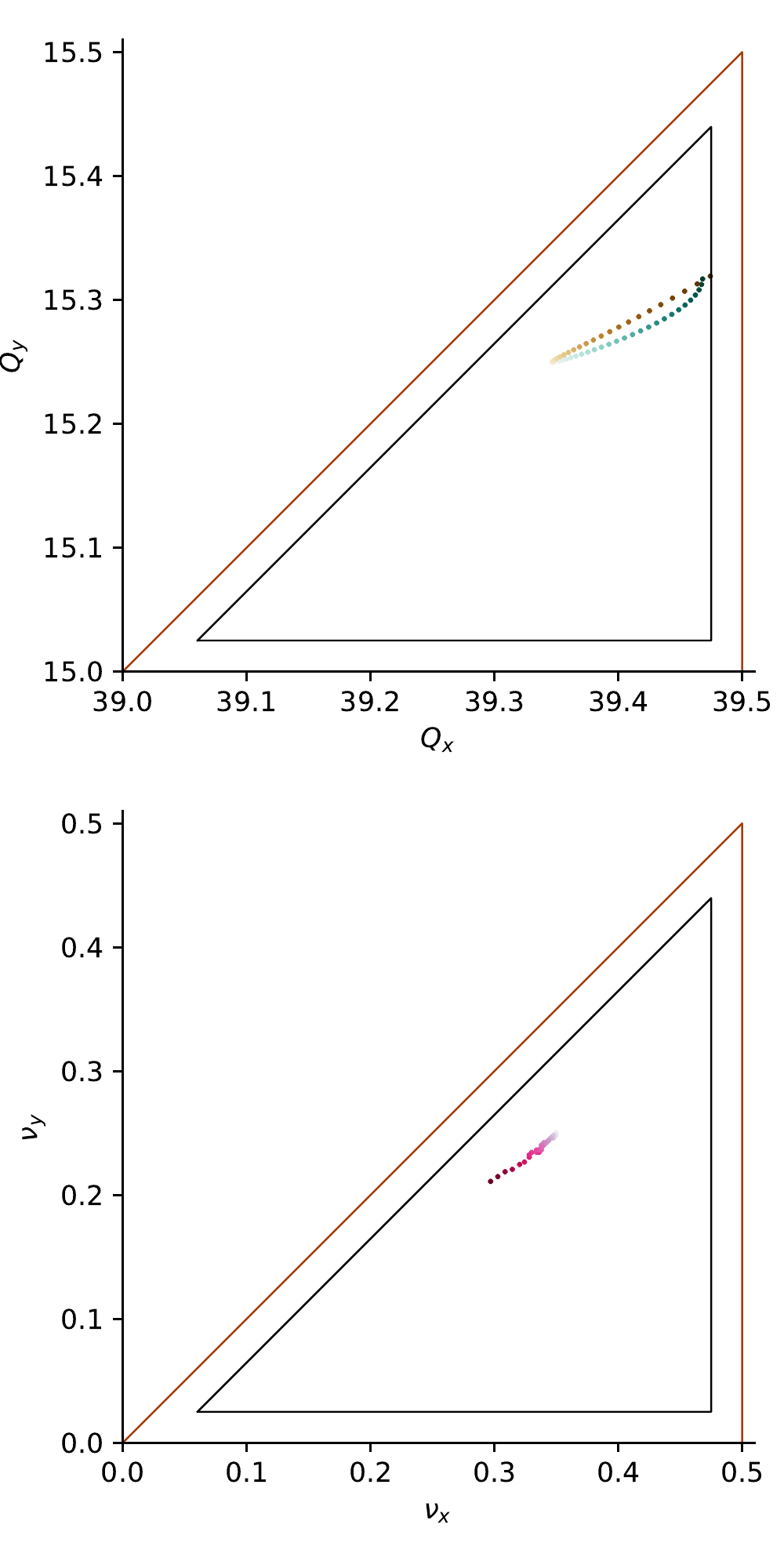}}
    \subfloat[\texttt{point-11} from Table~\ref{Tab:notable-points}.\label{fig:footprint-point11}]{
    \includegraphics[width=0.245\textwidth]{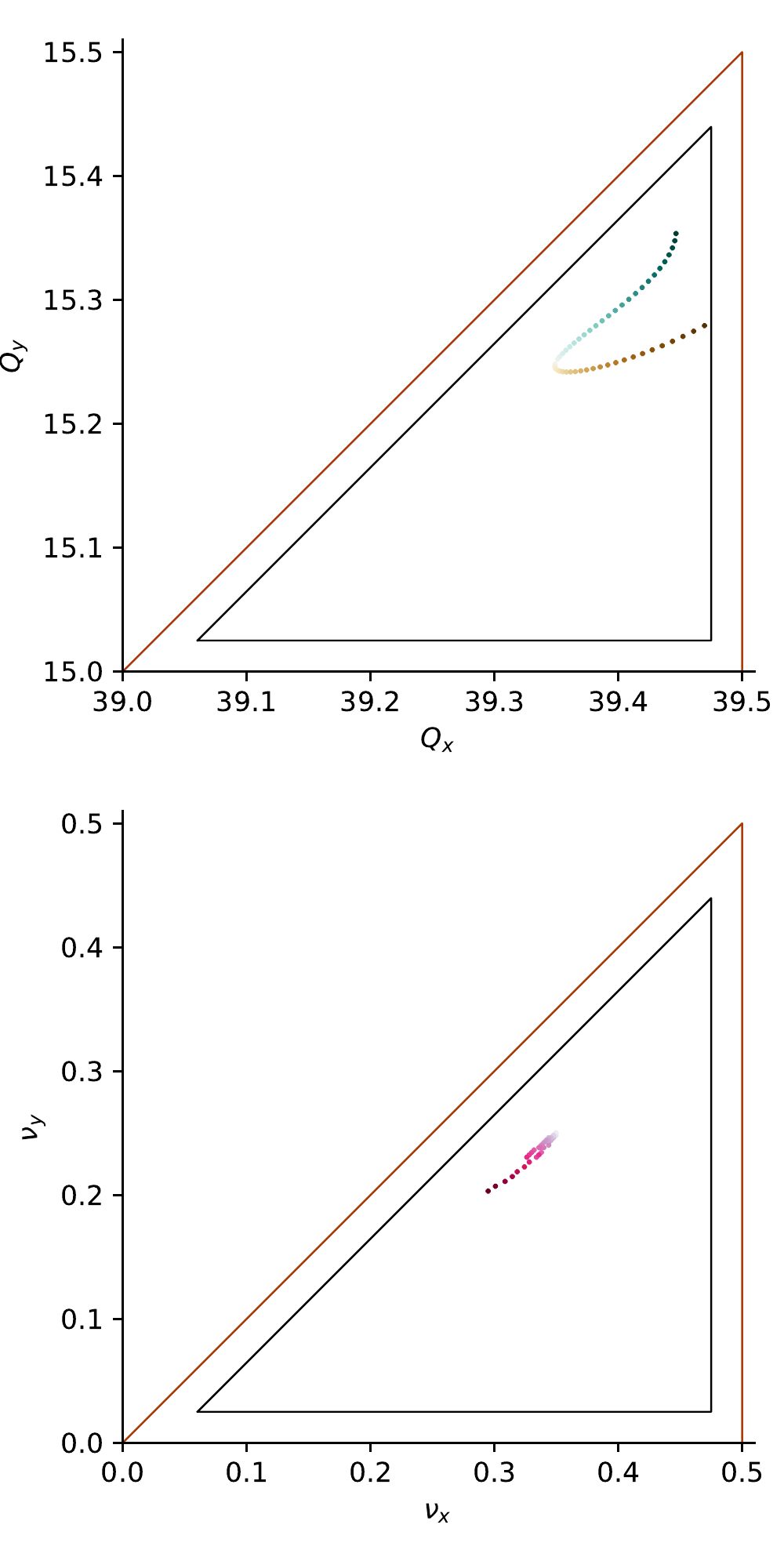}}
    \caption{Chromatic tune footprint in the range $\delta\in[-0.05,0.05]$ (top row) and amplitude-dependent tune footprint (bottom row) for the solution candidates: the manually obtained solution (referred to as the `design solution', 1st column), \texttt{point-3} from Table~\ref{Tab:b046_opt-pilot} (2nd column), the design point from Eq.~\eqref{eq:point-12} (3rd column) and \texttt{point-11} from Table~\ref{Tab:notable-points} (4th column).
    The 2nd order resonances are shown as red and the black triangle includes the additional margin around the resonance lines.
    \label{fig:footprints}}
\end{figure*}

The chromatic tune footprint (section~\ref{sec:ctfp}) and amplitude-dependent tune footprint (section~\ref{sec:adts}) for the three new solution candidates and the design solution are shown in Fig.~\ref{fig:footprints}. In each sub-plot, the outer triangle (red) is the one formed by three intersecting 2nd order resonances around the on-momentum tune (see Eq.~\eqref{eq:triangle}). The inner triangle (black) includes the margin from section~\ref{sec:ctfp}.
For all of the candidate solutions, $F_{4,5} = \texttt{unstable}_{-,+} = 0$ (see Eq.~\eqref{eq:unstable}) and the second constraint in Eq.~\eqref{eq:CMOOP-2} is satisfied, i.e., $\texttt{ctfp} + \texttt{adts} = 0$ (see Eqs.~\eqref{eq:ctfp} and \eqref{eq:adts}), so all footprints are located inside the inner triangle.

\section{Conclusions\label{sec:conclusions}}

In this paper a multi-objective genetic algorithm is used to find a good dynamic aperture and energy acceptance for the Swiss Light Source upgrade. 
To speed up this expensive computation, artificial neural network surrogate models are used in the optimization. 
Compared to a massively parallel implementation of a multi-objective genetic algorithm, the new optimization method results in an order of magnitude speedup. At the same time, the solution quality is preserved. In particular, tens of the design points in the last generation are better than the design solution in all of the considered objective functions. 

The new, faster method makes it possible to include more design parameters in the optimization problem, such as the octupole strengths, which could further improve the solution quality. Furthermore, it allows for the inclusion of a more accurate and more expensive model, e.g., a model which includes nonlinear synchrotron oscillation.
In this work the focus is on the lattice for the Swiss Light Source upgrade, but an analogous procedure could easily be used for a different lattice or a different machine.

\section{Acknowledgements}
M.~Aiba provided the manually obtained solution.
J.~Kallestrup found useful reference suggestions.
R.~Bellotti helped with \texttt{TensorFlow} and with proofreading a part of the manuscript. M.~Zacharias helped with \texttt{TensorFlow}.
T.~Schietinger proofread the manuscript.

\end{document}